\documentclass{aa}

\usepackage{rotating}
\usepackage{graphicx}
\usepackage{times}

\begin{document}
\title{Quasi-simultaneous {\it XMM-Newton} and VLA observation of the 
non-thermal radio emitter HD\,168112 (O5.5III(f$^+$))\thanks{Based on observations with 
XMM-Newton, an ESA Science Mission with instruments and contributions directly 
funded by ESA Member states and the USA (NASA). Also based on observations 
collected with the VLA, an instrument of the National Radio Astronomy Observatory,  
which is a facility of the National Science Foundation operated by Associated Universities, Inc. Optical data were collected at the European 
Southern Observatory (La Silla, Chile), and at the Observatorio Astron\'omico 
Nacional of San Pedro M\'artir (Mexico)}}

\author{M.\,De Becker\inst{1} \and G.\,Rauw\inst{1}\thanks{Research Associate 
FNRS (Belgium)} \and R.\,Blomme\inst{2} \and W.L.\,Waldron\inst{3} \and 
H.\,Sana\inst{1}\thanks{Research Fellow FNRS (Belgium)} \and 
J.M.\,Pittard\inst{4} \and P.\,Eenens\inst{1,5} \and I.R.\,Stevens\inst{6} \and
M.C.\,Runacres\inst{2} \and S.\,Van Loo\inst{2} \and A.M.T.\,Pollock\inst{7}} 

\institute{Institut d'Astrophysique, Universit\'e de Li\`ege, All\'ee du 6 
Ao\^ut, B\^at B5c, B-4000 Li\`ege (Sart Tilman), Belgium \and
Royal Observatory of Belgium, Avenue Circulaire 3, 1180 Brussels, Belgium \and
L-3 Communications Government Services, Inc., 1801 McCormick Drive, Suite 170, Largo, MD 20774, USA \and
Department of Physics \& Astronomy, University of Leeds, Leeds LS2 9JT, UK \and
Departamento de Astronomia, Universidad de Guanajuato, Apartado 144, 36000 
Guanajuato, GTO, Mexico \and
School of Physics \& Astronomy, University of Birmingham, Edgbaston 
Birmingham B15 2TT, UK \and
ESA/Vilspa, Apartado 50727, 28080 Madrid, Spain}

\date{Received date / Accepted date}
\authorrunning{M.\,De Becker et al.}
\titlerunning{An {\it XMM-Newton} and VLA observation of HD\,168112}

\abstract{We report the results of a multiwavelength study of the non-thermal radio emitter HD\,168112 (O5.5III(f$^+$)). The detailed analysis of two quasi-simultaneous {\it XMM-Newton} and VLA observations reveals strong variability of this star both in the X-ray and radio ranges. The X-ray observations separated by five months reveal a {\em decrease} of the X-ray flux of $\sim$\,30\,\%. The radio emission on the other hand {\em increases} by a factor 5--7 between the two observations obtained roughly simultaneously with the {\it XMM-Newton} pointings. The X-ray data reveal a hard emission that is most likely produced by a thermal plasma at kT\,$\sim$\,2--3 keV while the VLA data confirm the non-thermal status of this star in the radio waveband. Comparison with archive X-ray and radio data confirms the variability of this source in both wavelength ranges over a yet ill defined time scale. The properties of HD\,168112 in the X-ray and radio domain point towards a binary system with a significant eccentricity and an orbital period of a few years. However, our optical spectra reveal no significant changes of the star's radial velocity suggesting that if HD\,168112 is indeed a binary, it must be seen under a fairly low inclination. 
}
\maketitle

\section{Introduction}

Over the last two decades, X-ray and radio emission of O-type and Wolf-Rayet stars have been intensively studied (e.g., Raassen et al.\,\cite{Raassen}, Table\,9; Cappa et al.\,\cite{Cappa}). Among the early-type stars detected at radio wavelengths, a subset display unusually high radio flux levels with spectral shapes that deviate significantly from the expectations for thermal free-free emission (Bieging et al.\,\cite{BAC}; Williams \cite{Wil}). This feature is attributed to non-thermal emission, probably synchrotron radiation (White \cite{Wh}). Such an emission from stellar winds allows us to state that a magnetic field and a population of relativistic electrons exist in the radio emission region. Due to the large opacity of the stellar wind at radio wavelengths, the detected non-thermal free-free emission must arise far away from the stellar surface ($\sim$\,100\,R$_*$), outside the radio photosphere.

Two scenarios have been proposed where relativistic electrons are accelerated through the first order Fermi mechanism occurring in shocks. In the first one, the star is single and electrons are accelerated by shocks distributed across the stellar wind (Chen \& White \cite{CW} and references therein), and arising from the line-driving instability (also known as the deshadowing instability, Feldmeier\,\cite{feldth}). In the second scenario, the star is a binary and the shock responsible for the acceleration of electrons to relativistic 
velocities results from the collision of the stellar winds of the binary components (van der Hucht et al.\,\cite{vdh1}; Eichler \& Usov \cite{EU}; Dougherty et al. \cite{DP}).\\ 

The existence of a population of relativistic electrons implies that {\em high energy} non-thermal emission could be produced within the stellar winds and/or a within a wind-wind binary collision zone (Pollock \cite{Pol}, Chen \& White \cite{CW}). The idea is that Fermi accelerated relativistic electrons could be thermalized by the strong UV flux from the hot star through Inverse Compton (IC) scattering thereby generating non-thermal X-ray and $\gamma$-ray emission. Following this model, X-ray spectra of such objects should display the signature of IC scattering emission through a hard power law tail in the high energy spectrum (above $\sim$ 3 keV).\\

Dougherty \& Williams (\cite{DW}) showed that out of nine non-thermal radio emitting Wolf-Rayet (WR) stars, seven are visual or spectroscopic binaries. Although the fraction of binaries among the non-thermal emitting O-stars is much less constrained, one may wonder whether binarity is indeed a necessary condition for the occurrence of non-thermal emission and whether these stars exhibit also non-thermal X-ray emission. In an attempt to answer this question, Rauw et al.\,(\cite{9sgr}) investigated the case of the O4((f$^+$)) star 9\,Sgr (HD\,164794). This presumably single star is indeed a well-known non-thermal radio emitter, and consequently a good candidate to search for a high energy non-thermal counterpart. While 9\,Sgr indeed exhibits a hard emission tail in its X-ray spectrum, the thermal or non-thermal nature of this tail could not be established. On the other hand, the optical spectra analysed by Rauw et al.\ (\cite{9sgr}) show some clues that 9\,Sgr might be a long period binary.

In the framework of the same campaign, we investigated the case of another presumably single non-thermal radio emitter: HD\,168112. In this purpose, we obtained two sets of quasi simultaneous observations (March/April and September 2002) with the {\it XMM-Newton} satellite and the NRAO Very Large Array (VLA) radio observatory, as well as optical data spread over two years. This rather luminous star (m$_\mathrm{V}$ = 8.55) lies inside NGC\,6604, which is a fairly compact open cluster lying at the core of the HII region S54 (Georgelin et al.\,\cite{Geor}). Conti \& Ebbets (\cite{CE}) derived an O5.5((f)) spectral type for HD\,168112, while Walborn (\cite{Wal}) assigned an O5\,III type. This star is known as a non-thermal radio emitter, with a negative radio spectral index, and a flux level at 6\,cm ranging between 1.3 and 5.4\,mJy (Bieging et al.\,\cite{BAC}).\\

This paper is devoted to HD\,168112. The analysis of the observations of other members of the NGC\,6604 cluster, among which the luminous multiple star HD\,167971 (O5/8V + O5/8V (+ O8I), Leitherer et al.\,\cite{Lei2}), will be addressed in a forthcoming paper. In Sect.\,\ref{sect_obs}, we describe the {\it XMM-Newton}, VLA, and optical observations. Sections \ref{sect_epic1} and \ref{var} are respectively devoted to the analysis of the X-ray spectra and the X-ray variability between our two {\it XMM-Newton} observations. Section \ref{former} briefly discusses former observations of HD\,168112 in the X-ray domain whilst Sects.\,\ref{sect_radio} and \ref{sect_opt} deal with the analysis of the radio and optical data. The discussion and conclusions are respectively presented in Sects.\,\ref{disc} and \ref{concl}.\\ 

\section{Observations \label{sect_obs}}
\subsection{X-ray data}
NGC\,6604 was observed with {\it XMM-Newton} (Jansen et al.\,\cite{xmm}) during 
revolution 426, on April 6, 2002 (Obs.\,ID 0008820301, JD\,2452372.477 -- 
.637), and revolution 504, on September 9 (Obs.\,ID 0008820601, 
JD\,2452526.694 -- .868). All three EPIC instruments (MOS1, MOS2, and pn) were 
operated in the full frame mode (Turner et al.\,\cite{mos}, Str\"uder et 
al.\,\cite{pn}), and used the thick filter to reject optical light. We did not 
obtain any RGS data because HD\,167971 and HD\,168112 were observed off-axis to 
make sure that they appear simultaneously in the EPIC field of view. {\bf Because of the brightness of the sources in the field, the Optical Monitor was closed during the observations.}\\ 

The data reduction was performed with version 5.3.1 of the {\it XMM} 
Science Analysis System (SAS). The raw files were processed through the `emproc' and `epproc' tasks respectively for EPIC-MOS and EPIC-pn. For EPIC-MOS data analysis, we considered only events with pattern 0 -- 12, and no significant pile-up was found (pattern 26 -- 31). The case of EPIC-pn was more difficult. Indeed, we first applied too severe screening criteria (pattern 0 -- 4, flag = 0), such that bad columns and bad pixels disappeared, as well as pixels along the CCD gaps. Unfortunately, because of the position of HD\,168112, the source region is crossed by a gap, and a bad column. As a consequence, only a few percent of the star events remained, which was not sufficient for a reliable analysis to be performed. We therefore relaxed the selection criteria, and selected events with pattern 0 -- 12. Consequently, we have to consider our EPIC-pn spectra of HD\,168112 with some caution. Nevertheless, the pn data will be compared to the MOS data in the following sections. Finally, response matrices and ancillary files were generated through the `rmfgen' and `arfgen' tasks respectively.

\begin{figure}
\begin{center}
\resizebox{8.5cm}{7.0cm}{\includegraphics{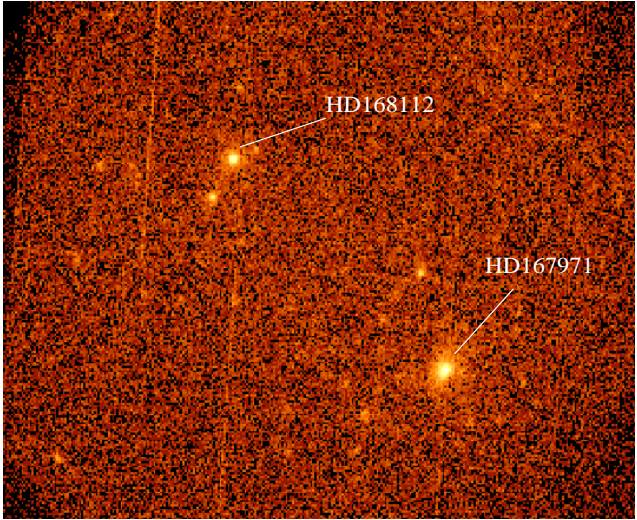}}
\caption{Combined EPIC (MOS1, MOS2 and pn) image obtained for the first 
exposure (April 2002). The brightest source is the eclipsing binary 
HD\,167971. HD\,168112 is the second brightest source of the field. Other members of the NGC\,6604 open cluster appear also as X-ray emitters. The width of the field is about 30\,arcmin. The North is up, and the East is on the left.\label{field}}
\end{center}
\end{figure}

The combined EPIC image in Fig.\,\ref{field} shows that the two brightest 
sources are, in order of increasing luminosity, HD\,168112 and HD\,167971. Other point sources, presumably belonging to the NGC\,6604 open cluster also appear in the field. As already mentioned, the analysis of these sources is postponed to a forthcoming study.\\

\subsection{Radio data}

Table~\ref{table VLA observing log} presents the log of the NRAO VLA observations. The two observing runs were performed close to the {\em XMM-Newton} ones: the first VLA observation took place 14 days before the corresponding {\it XMM-Newton} pointing and the second VLA observation was obtained 2 days after the second X-ray run. For both epochs, data were obtained at 3.6\,cm (X-band), 6\,cm (C-band) and 20\,cm (L-band). For the second epoch we also observed at 18\,cm. Fluxes at each wavelength were calibrated on the flux calibrator 3C48 = 0137+331 (J2000).
The fluxes assigned to the flux calibrator are given in Table~\ref{table fluxes flux calibrator}. Each observation of HD\,168112 was preceeded and followed by an observation of the phase calibrator 1832-105 (J2000) for the 3.6 and 6\,cm bands, and of 1834-126 (J2000) for the 18 and 20\,cm ones. All observations were made in two sidebands (denoted IF1 and IF2), each of which has a bandwidth of 50 MHz.

The data reduction was done using the Astronomical Image Processing System
(AIPS), developed by NRAO. The observed visibilities of the calibrators were used to determine the instrumental gains. These were then interpolated in time and applied to the HD\,168112 visibilities. The Fourier transform of the calibrated visibilities (using robust uniform weighting -- see Briggs~\cite{Briggs95}) resulted in an intensity map, which was then cleaned to remove the effect of the beam.

\begin{table}
\caption{Log of VLA observations. The date, programme and configuration
are listed, as well as the wavelength observed and the integration time
on HD\,168112 (in minutes). {\bf Fluxes are given in Table\,\ref{table VLA flux results}.}}
\label{table VLA observing log}
\begin{center}
\begin{tabular}{lllr@{.}lr}
\hline \hline
\multicolumn{1}{c}{Date} & \multicolumn{1}{c}{Progr.} & 
   \multicolumn{1}{c}{Config.} & \multicolumn{2}{c}{$\lambda$} & 
   \multicolumn{1}{c}{Int. time} \\
 & & & \multicolumn{2}{c}{(cm)} & \multicolumn{1}{c}{(min.)} \\
\hline
2002-Mar-24 & AB1048   & \multicolumn{1}{c}{A}        &  3&6 & 11.0 \\
\multicolumn{2}{l}{(JD=2452358.089-.148)} &           &  6&  & 10.8 \\
       &                                &             & 20&  &  5.0 \\
2002-Sep-11 & AB1065    & \multicolumn{1}{c}{CnB$^a$} &  3&6 &  8.5 \\
\multicolumn{2}{l}{(JD=2452528.662-.723)} &           &  6&  &  8.3 \\
       &                                &             & 18&  &  5.0 \\
       &                                &             & 20&  &  4.7 \\
\hline
\end{tabular}
\end{center}
$^a$: North arm is in B configuration, East and West arm are in 
C configuration but with 3 antennas still
in B-type positions.\\
\end{table}

\begin{table}
\caption{Fluxes of the flux calibrator 3C48, based
on the 1995.2 VLA coefficients (Perley \& Taylor 1999). The two values
for each entry correspond to the two 50 MHz sidebands (IF1 and IF2).}
\label{table fluxes flux calibrator}
\begin{center}
\begin{tabular}{r@{.}lrrrrr}
\hline \hline
\multicolumn{2}{c}{$\lambda$} & \multicolumn{2}{c}{Frequency (GHz)} & 
   \multicolumn{2}{c}{Flux (Jy)} \\
\multicolumn{2}{c}{(cm)} & \multicolumn{1}{c}{IF1} & \multicolumn{1}{c}{IF2} & 
   \multicolumn{1}{c}{IF1} & \multicolumn{1}{c}{IF2} \\
\hline
 3&6 & 8.485 & 8.435 &  3.1449 &  3.1637 \\
 6&  & 4.885 & 4.835 &  5.4054 &  5.4585 \\
18&  & 1.665 & 1.635 & 13.9788 & 14.1852 \\
20&  & 1.465 & 1.385 & 15.4935 & 16.1955 \\
\hline
\end{tabular}
\end{center}
\end{table}

As HD\,168112 is close to the Galactic plane, background structure is expected to be present at radio wavelengths. Only the lower-resolution 18 and 20\,cm maps taken on the second epoch show some background. To eliminate this, we dropped all observational data taken on the shortest baselines for these two observations.

\subsection{Optical data}
Four echelle spectra of HD\,168112 were gathered with the Fiber-fed Extended 
Range Optical Spectrograph (FEROS, Kaufer et al.\,\cite{Kaufer}). In May 2001 and April 2002 the spectrograph was attached to the ESO 1.52\,m telescope at La 
Silla, while in May 2003, it was used at the 2.2\,m ESO/MPE telescope at La 
Silla. The exposure times varied between 25 and 45 minutes. The spectral 
resolving power of the FEROS instrument is 48\,000. The detector was an EEV CCD 
with $2048 \times 4096$ pixels of $15\,\mu$m$ \times 15$\,$\mu$m. We used an 
improved version of the FEROS context within the {\sc midas} package provided by ESO to reduce the data.

A series of spectra of HD\,168112 were also obtained in April 2003 with the 
echelle spectrograph at the 2\,m telescope of the Observatorio Astron\'omico 
Nacional of San Pedro M\'artir (SPM). The detector was a Site CCD with 
1024$\times$1024 pixels of 24\,$\mu$m squared. The slit width was set to 
150\,$\mu$m corresponding to 2\,\arcsec\ on the sky. The data were reduced using the {\sc midas} echelle package and the individual wavelength-calibrated echelle orders were rectified using carefully selected continuum windows.

\section{The EPIC spectrum of HD\,168112 \label{sect_epic1}}
\subsection{Dealing with high background level episodes \label{sect_flare}}

\begin{figure}
\begin{center}
\resizebox{8.5cm}{5.0cm}{\includegraphics{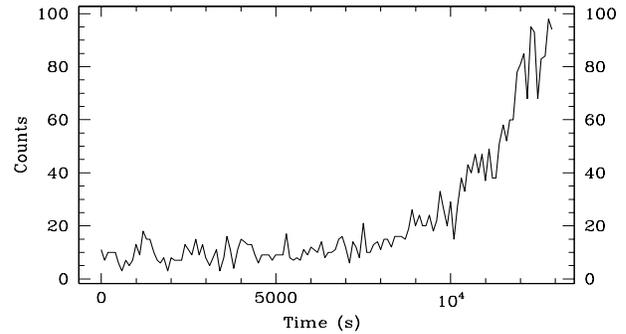}}
\caption{Raw light curve for events with PI $>$ 10\,000 for MOS1 during the first exposure, obtained with a time bin of 100\,s. The second part of the exposure (after time 9000\,s) is strongly affected by the flare. 
\label{flare}}
\end{center}
\end{figure}

A raw light curve extracted at very high energies (Pulse Invariant (PI) channel 
numbers $>$ 10\,000) revealed a strong soft proton flare (Lumb \cite{lumb}) during the second half of the first exposure as shown in Fig.\,\ref{flare}.

\begin{figure}
\begin{center}
\resizebox{8.5cm}{5.0cm}{\includegraphics{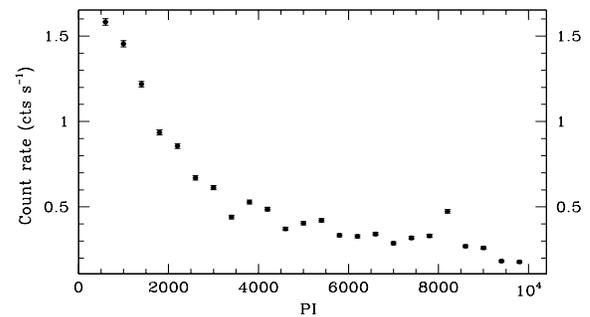}}
\caption{Total count rate over the 350\,arcmin$^2$ background area (see text) versus PI during the soft proton flare of the first exposure with PI $\in$ [400:10\,000] for EPIC-pn. The general shape of the curve is the same as for the mean background before the flare but displays an intensity level that is a factor about 2.5 times higher. The bump near PI 8000 is produced by fluorescent lines generated by the interaction of charged particles with the body of the pn detector (Lumb\,\cite{lumb}). The error bars represent the 1-$\sigma$ standard deviation.\label{flare2}}
\end{center}
\end{figure}

The level of the overall background during the flare was estimated following the same procedure as described by De Becker et al.\ (\cite{DeB}) in the case of the {\it XMM-Newton} observation of the colliding wind binary HD\,159176. We selected events occuring on the whole EPIC-pn detector with boxes excluding point sources, gaps, and bad columns, leading to a total extraction area of about 350 arcmin$^2$ ($\sim$ half of the EPIC-pn field area). This event list was then used to obtain light curves over intervals 400 PI channels wide, with PI $\in$ [400:10\,000]. This operation was performed both on the low background and on the flare parts of the first observation to get a rough estimate of the overall background as a function of the energy. A comparison of these two results reveals that the mean overall background level during the flare is about 2.5 times higher than the mean level during the quiet background period. The overall background count rate is maximum at lower energies and decreases rapidly as PI increases, as shown in Fig.\,\ref{flare2}. After applying the suitable area scaling factor, we compared the count rates of the overall background and of the background corrected source region in each PI interval. It appeared that the background level during the flare is of the same order of magnitude as the source at energies below 5 keV.\\  

We rejected the bad time interval by applying a threshold of 0.16\,cts\,s$^{-1}$ in the case of EPIC-MOS, and 1.10\,cts\,s$^{-1}$ for EPIC-pn to obtain the 
filtered event list. The first exposure ($\sim$ 13 ks) was strongly affected by the flare and only about 7 ks were usable for data analysis after filtering. The whole second exposure ($\sim$ 14 ks) was free from flaring events and no time filtering was considered. To check for the impact of this flare on our spectral analysis of the April 2002 data, we extracted spectra from our filtered and unfiltered event lists for the three EPIC instruments. Both sets of spectra were fitted with some of the models described in Sect.\,\ref{sect_fit}. The results we obtained were identical within the error bars for both spectral sets. This is because the background subtraction (which is a part of our reduction procedure) cancels the effect of the soft proton flare. As a consequence, we consider that the spectral analysis, at least for a source as bright as HD\,168112, can be performed on the whole data set. In the following, all fitting results will refer to the whole exposure for the April observation, without any filtering for soft proton contaminated time intervals. As a consequence, the effective exposure time is very close to the performed duration of the exposure, as shown in Table\,\ref{expt}.\\

\begin{table}
\caption{Performed duration and effective exposure time of both observations 
(April and September 2002). The difference between the performed and effective 
duration of the exposures is only due to overheads. \label{expt}}
\begin{center}
\begin{tabular}{ccccc}
\hline \hline
	& \multicolumn{2}{c}{Obs.\,1} & \multicolumn{2}{c}{Obs.\,2} \\
\cline{2-3}\cline{4-5}
Instrument	& Performed & Effective & Performed & Effective \\
	& duration & exposure & duration & exposure \\
	& (s) & (s) & (s) & (s) \\
\hline
MOS1 & 13\,122 & 12855 & 13\,667 & 13\,402 \\
MOS2 & 13\,122 & 12896 & 13\,667 & 13\,406 \\
pn & 10\,546 & 9525 & 12\,046 & 10\,771 \\
\hline
\end{tabular}
\end{center}
\end{table}

\subsection{Spectral analysis \label{sect_fit}}

The HD\,168112 EPIC spectra were extracted within a circular region of radius 60 arcsec. The actual background spectrum was derived from an annulus centered on the source (the external radius is about 85 arcsec), excluding its intersection with a circular region centered on a point source (RA=18:18:44.3 and DEC=$-$12:07:51.5, Equinox 2000.0). The figure provided in gif format shows the source and background regions selected in the case of EPIC-MOS1 data. In the EPIC-pn case, a rectangular region was also considered to exclude the gap between CCDs which crosses the source and/or background regions.\\

\begin{figure*}[htb]
\begin{center}
\resizebox{17.0cm}{6.0cm}{\includegraphics{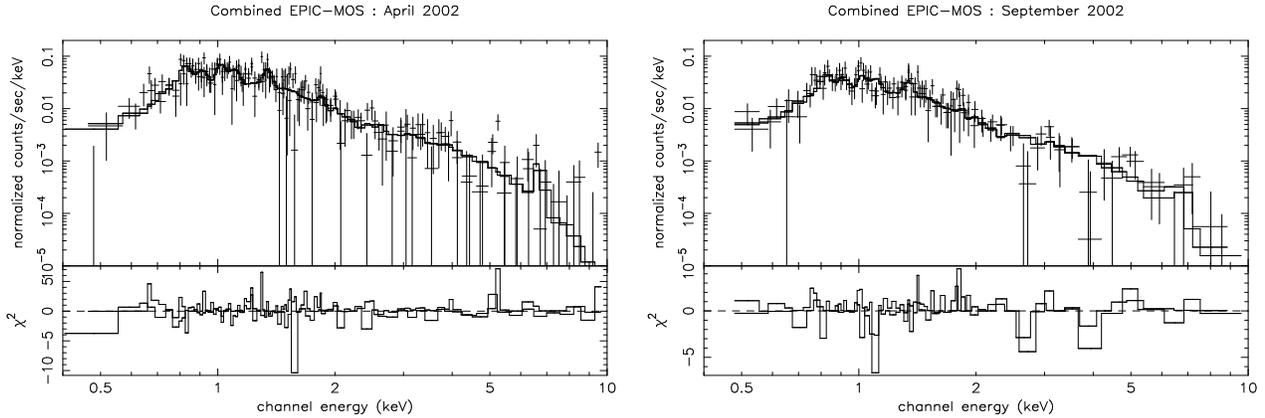}}
\caption{Combined EPIC-MOS1 and EPIC-MOS2 spectra of HD\,168112 fitted with a 
{\tt wabs$_\mathrm{ISM}$*wind*(2-T mekal)} model between 0.4 and 10.0 keV. The ISM absorption component is frozen at 0.58\,$\times$\,10$^{22}$\,cm$^{-2}$. The two model-lines appearing in the upper part are due to the fact that the response matrices of the two instruments are different. Left panel: observation 1. Right panel: observation 2.\label{spmos}}
\end{center}
\end{figure*}

EPIC-MOS and EPIC-pn spectra were rebinned to get a minimum of 9 and 16 counts respectively in each energy channel. We fitted spectra within {\sc xspec} (release 11.1.0) using different components including optically thin thermal plasma {\tt mekal} models (Mewe et al.\,\cite{mewe}, Kaastra \cite{ka}), as well as power laws. On the one hand, the choice of the {\tt mekal} model was motivated by the fact that massive stars are known to display rather soft thermal spectra, as confirmed for example by RGS spectra of $\zeta$\,Pup (Kahn et al.\,\cite{Kahn}) or HD\,159176 (De Becker et al.\,\cite{DeB}). This thermal emission is nowadays attributed to shock heated plasma resulting either from instability-induced shocks distributed throughout the stellar wind (Feldmeier et al.\,\cite{Feld}) or from a colliding wind interaction in a binary system (Stevens et al.\,\cite{SBP}). On the other hand, as HD\,168112 is known as a non-thermal radio emitter, one should also consider the possibility of a non-thermal X-ray counterpart, which would dominate the spectrum at hard energies and should then be modelled by a power law component (Pollock \cite{Pol}; Chen \& White \cite{CW}).\\

The quality of our fittings was estimated through the $\chi^2$ minimization, and the estimation of the model parameters was obtained on the basis of the resulting best fit. However, one could ask whether the $\chi^2$ statistic is applicable to the case of bins containing a small amount of counts. For this reason, we checked the consistency of our results with those obtained with the C-statistic (Cash \cite{cash}). For a specific spectral model, this method provides a alternative approach for parameter estimation in the case of small count numbers. Another technique which was considered is a $\chi^2$ statistic with a Churazov weighting, also well designed for small count numbers (Churazov et al.\,\cite{chur}). As a result, we obtained parameter values which are in agreement with the values given in Tables\,\ref{fitwTT} and \ref{fitwTP} within the error bars. For this reason, only the standard $\chi^2$ technique is considered in Sect.\,\ref{fit}.\\ 

\subsubsection{Absorption columns}
Different combinations of the above emission components (mekal + power) were tried, to fit the whole X-ray emission within the EPIC bandpass, i.e.\ between about 0.4 and 10.0 keV. We also included in our fittings the absorption due to both the interstellar medium (ISM) and the wind 
material.\\

The ISM absorption column was frozen at 0.58\,$\times$\,10$^{22}$\,cm$^{-2}$. This value 
was obtained assuming N$_\mathrm{H}$ = 
5.8\,$\times$\,10$^{21}$\,$\times$\,E(B--V)\,cm$^{-2}$ as given by Bohlin et 
al.\,(\cite{Boh}).  The colour excess was evaluated from an observed 
(B--V) of +0.69 (Chlebowski et al.\,\cite{Chle}), and an intrinsic colour 
(B--V)$_\mathrm{o}$ of --0.309 (following e.g. Mihalas \& 
Binney\,(\cite{MB}) for an O5 star).

The local absorption column, related to the wind material, has been considered 
following two approaches. The first one consisted in considering a cold neutral 
absorption column, as for the case of the interstellar column. In the second 
approach, we treated the X-ray absorption by the wind in a more realistic way 
using a stellar wind model (Waldron et al.\,\cite{Waldron+al98}). The purpose of this model is to deal with the absorption by warm ionized material, taking into 
account the fact that ionization edges of ionized elements are significantly 
shifted to higher energies as compared to those of neutral elements. The wind 
column density above a given radius is calculated assuming a spherically symmetric wind and a standard $\beta$ velocity law. Cross sections for such a wind model differ essentially from those of cold material below about 1 keV. The stellar parameters used to model the wind opacity of HD\,168112 are given in 
Table\,\ref{param}. On the basis of these opacities, a FITS table was 
generated in a format suitable to be used within {\sc xspec} as a multiplicative model. The only free parameter of this model is the wind column density 
(N$_\mathrm{W}$). In the remaining of the text, this wind absorption model will 
be referred to as the {\tt wind} component. For both cold and warm absorption 
columns, solar abundances are assumed.\\

\begin{table}
\caption{Stellar parameters of HD\,168112 taken from 
Leitherer\,(\cite{Leitherer88})\label{param}}
\begin{center}
\begin{tabular}{lc}
\hline \hline
M\,(M$_{\odot}$) & 70 \\
R$_*$\,(R$_{\odot}$) & 16 \\
v$_{\infty}$\,(km\,s$^{-1}$) & 3250 \\
${\dot{\it {\rm M}}}$\,(M$_{\odot}$\,yr$^{-1}$) & 2.5\,$\times$\,10$^{-6}$ \\
L$_\mathrm{bol}$\,(erg\,s$^{-1}$) & 3.0\,$\times$\,10$^{39}$ \\
d\,(kpc) & 2 \\
\hline
\end{tabular}
\end{center}
\end{table}

Because of its more realistic treatment of wind absorption, emphasis will be 
put on the results obtained with the ionized material absorption column. 
Nevertheless both approaches will be discussed in the next section.

\subsubsection{Spectral fittings\label{fit}}

We first tried to fit both individual and combined EPIC-MOS spectra, with a cold local absorption component. A single temperature model appeared to be ineffective to fit data above about 3 keV ($\chi_\nu^2 \geq$ 1.3). Better results were achieved with a two-temperature model. The temperatures, consistent from one exposure to the other, are of the order of 0.2 keV and 2$-$3 keV for EPIC-MOS data (about 2.3\,$\times$\,10$^{6}$\,K and 3\,$\times$\,10$^{7}$\,K respectively). Only the emission measures suffer a decrease between the two observations. For the second observation, we noted some inconsistencies between the fitting parameters of the MOS1 and MOS2 spectra, especially for the absorption component and the normalization parameter of the softest component. This problem occurred also for the fittings with the {\tt wind} model and it turned out that it was due to a local minimum in the parameter space of the MOS2 spectral fit. To avoid such problems and to improve the statistics of our fittings we shall therefore focus on combined MOS1 + MOS2 fittings.

As stated in Sect.\,\ref{sect_obs}, our EPIC-pn data are to be considered with some caution.We first fitted pn spectra under the same conditions as for 
EPIC-MOS data. The results reveal spectral properties very similar to the EPIC-MOS ones, consisting of a soft component of 0.2 keV and a hard thermal component of about 2 -- 2.5 keV. Because of the quality of the fit and its rather good consistency with EPIC-MOS results, we decided to consider EPIC-pn results as well in our overall discussion of the X-ray spectrum of HD\,168112.\\

\begin{table*}[ht]
\caption{Best-fit parameters for EPIC spectra of HD\,168112 in the case of a {\tt wabs$_\mathrm{ISM}$*wind*(2-T mekal)} model. Results are given for combined MOS, pn, and combined EPIC spectra. The upper and bottom parts of the table concern the first and second observation respectively. The ISM absorption component ({\tt wabs$_\mathrm{ISM}$}) is frozen at 0.58\,$\times$\,10$^{22}$\,cm$^{-2}$. The quoted circumstellar absorption column is ionized, with N$_\mathrm{W}$ expressed in cm$^{-2}$. The last column gives the observed flux between 0.4 and 10.0 keV for each fitting result. The normalization parameter (Norm) of the {\tt mekal} component is defined as $(10^{-14}/(4\,\pi\,D^2))\int{n_\mathrm{e}\,n_\mathrm{H}\,dV}$, where $D$, 
$n_\mathrm{e}$ and $n_\mathrm{H}$ are respectively the distance to the source (in cm), and the electron and hydrogen number densities (in cm$^{-3}$). The error bars represent the 1-$\sigma$ confidence interval.\label{fitwTT}}
\begin{center}
\begin{tabular}{cccccccc}
\hline
\hline
	& $\log{\mathrm{N}_\mathrm{W}}$ & kT$_1$ & Norm$_1$ & kT$_2$ & Norm$_2$ & 
$\chi^2_\nu$ & Obs.\,flux \\
	&  & (keV) &  & (keV) &  & d.o.f. & (erg\,cm$^{-2}$\,s$^{-1}$)\\
\hline
\hline
MOS1 + MOS2 & 21.90 & 0.29 & 5.36\,$\times$\,10$^{-3}$ & 2.72 & 3.33\,$\times$\,10$^{-4}$ & 0.98 & 
4.77\,$\times$\,10$^{-13}$\\ 
	& $\pm$ 0.07 & $\pm$ 0.02 & $\pm$ 2.93\,$\times$\,10$^{-3}$ & $\pm$ 0.62 & $\pm$ 
1.01\,$\times$\,10$^{-4}$ & 174 & \\
\hline
pn	& 21.89 & 0.27 & 5.59\,$\times$\,10$^{-3}$ & 2.00 & 4.77\,$\times$\,10$^{-4}$ & 1.07 & 
4.78\,$\times$\,10$^{-13}$\\ 
	& $\pm$ 0.07 & $\pm$ 0.02 & $\pm$ 3.29\,$\times$\,10$^{-3}$ & $\pm$ 0.33 & $\pm$ 
1.14\,$\times$\,10$^{-4}$ & 136 & \\
\hline
MOS1 + MOS2 & 21.90 & 0.28 & 5.57\,$\times$\,10$^{-3}$ & 2.29 & 4.06\,$\times$\,10$^{-4}$ & 1.01 & 
4.79\,$\times$\,10$^{-13}$ \\
+ pn & $\pm$ 0.05 & $\pm$ 0.02 & $\pm$ 2.25\,$\times$\,10$^{-3}$ & $\pm$ 0.31 & $\pm$ 0.76\,$\times$\,10$^{-4}$ & 315 &  \\
\hline
\hline
MOS1	& 21.88 & 0.26 & 5.14\,$\times$\,10$^{-3}$ & 2.36 & 2.93\,$\times$\,10$^{-4}$ & 0.90 & 
3.59\,$\times$\,10$^{-13}$\\ 
+ MOS2	& $\pm$ 0.08 & $\pm$ 0.02 & $\pm$ 3.44\,$\times$\,10$^{-3}$ & $\pm$ 0.49 & $\pm$ 
0.82\,$\times$\,10$^{-4}$ & 116 & \\
\hline
pn	& 21.96 & 0.25 & 6.68\,$\times$\,10$^{-3}$ & 2.16 & 2.97\,$\times$\,10$^{-4}$ & 1.01 & 
3.34\,$\times$\,10$^{-13}$\\ 
	& $\pm$ 0.06 & $\pm$ 0.01 & $\pm$ 3.06\,$\times$\,10$^{-3}$ & $\pm$ 0.42 & $\pm$ 0.83\,$\times$\,10$^{-4}$ & 89 & \\
\hline
MOS1 + MOS2 & 21.93 & 0.25 & 6.14\,$\times$\,10$^{-3}$ & 2.23 & 2.97\,$\times$\,10$^{-4}$ & 0.94 & 
3.44\,$\times$\,10$^{-13}$ \\
+ pn & $\pm$ 0.05 & $\pm$ 0.01 & $\pm$ 2.41\,$\times$\,10$^{-3}$ & $\pm$ 0.32 & $\pm$ 0.59\,$\times$\,10$^{-4}$ & 210 &  \\
\hline
\hline
\end{tabular}
\end{center}
\end{table*}

As a second step, we performed the same fittings but using a warm local absorption component. The results are summarized in Table\,\ref{fitwTT}. Fig.\,\ref{spmos} shows combined EPIC-MOS1 and EPIC-MOS2 spectra as obtained 
respectively for the two observations in the case of a two-component thermal model. The temperatures obtained are about 0.3 keV and 2 -- 3 keV (about 3.5\,$\times$\,10$^{6}$\,K and 3\,$\times$\,10$^{7}$\,K respectively), in rather good agreement with those obtained for the cold absorption model. As could be expected, the soft emission component is mostly affected by the change of the absorption model. We emphasize on the compatibility between the parameters of the combined EPIC-MOS and the combined EPIC-MOS/pn fittings.

We also tried to fit similar models but using two local absorption components, one for each individual {\tt mekal} component. However, it turned out that one of the absorption components could not be constrained by the fit. For this reason, we do not consider the results of this multiple absorption components model in our discussion.\\  

\begin{figure}[ht]
\begin{center}
\resizebox{8.5cm}{6.0cm}{\includegraphics{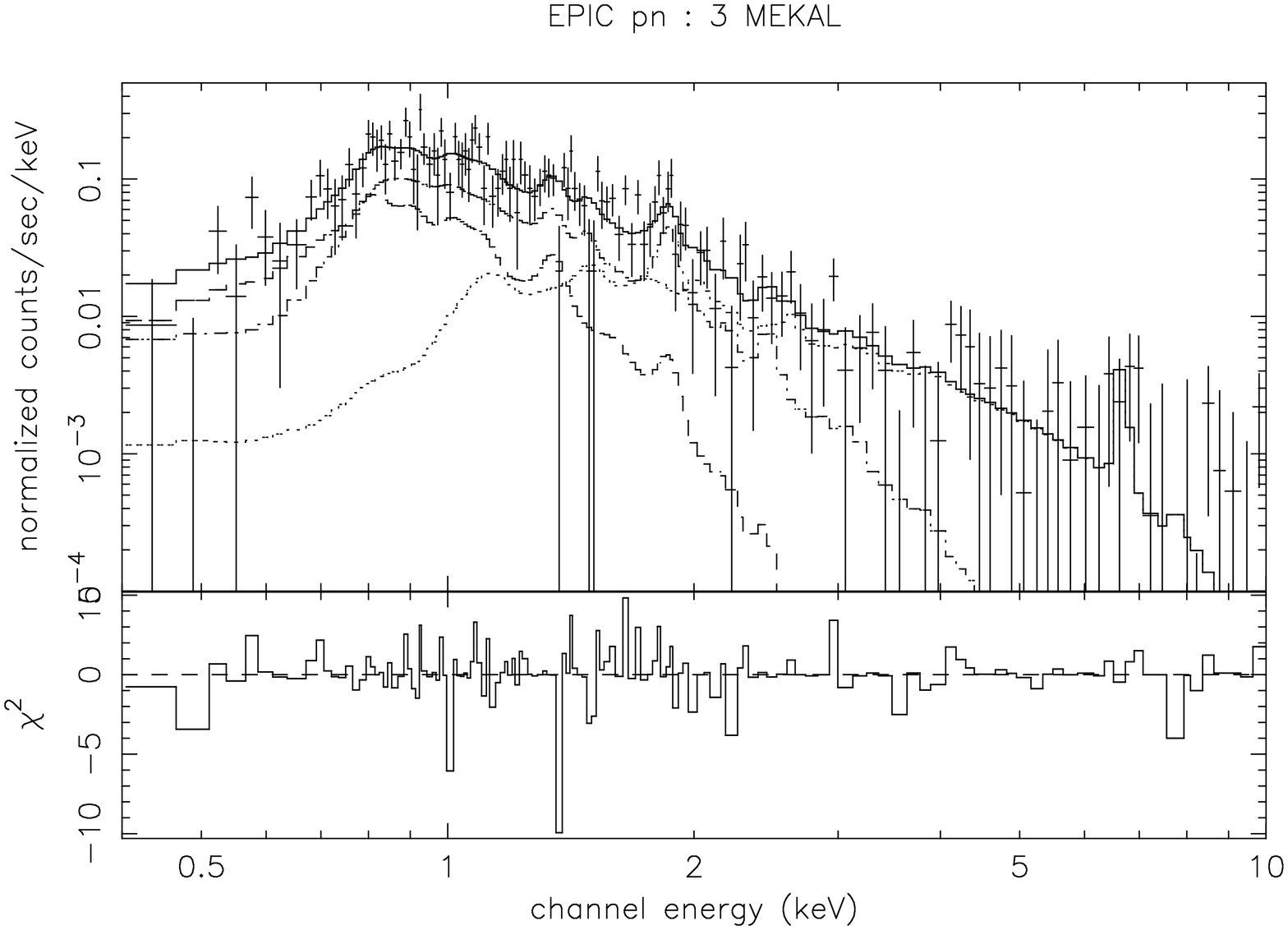}}
\caption{EPIC-pn spectrum of HD\,168112 fitted with a {\tt 
wabs$_\mathrm{ISM}$*wind*(3-T mekal)} model (solid line) between 0.4 and 10.0 keV for the first observation (April 2002). The ISM absorption component is frozen at 0.58\,$\times$\,10$^{22}$\,cm$^{-2}$. The three thermal components are displayed individually (dotted, dashed-dotted, and dashed lines). \label{pnfit}}
\end{center}
\end{figure}

Another uncertainty affects our analysis of the EPIC data with the models discussed hereabove. This issue concerns the fact that the fit results described are related to a local minimum. The other local minimum gave a temperature for the soft component of about 0.55 -- 0.60 keV. The quality of the fit was very similar to that of the previous one obtained with the same data sets. However, it appeared that in some cases this local minimum gave rise to inconsistent values for the warm local absorption component, with rather large error bars as compared to those quoted in Table\,\ref{fitwTT}. For this reason, only the results relevant to the local minimum described in Table\,\ref{fitwTT} are considered.

\begin{figure*}[htb]
\begin{center}
\resizebox{17.0cm}{5.5cm}{\includegraphics{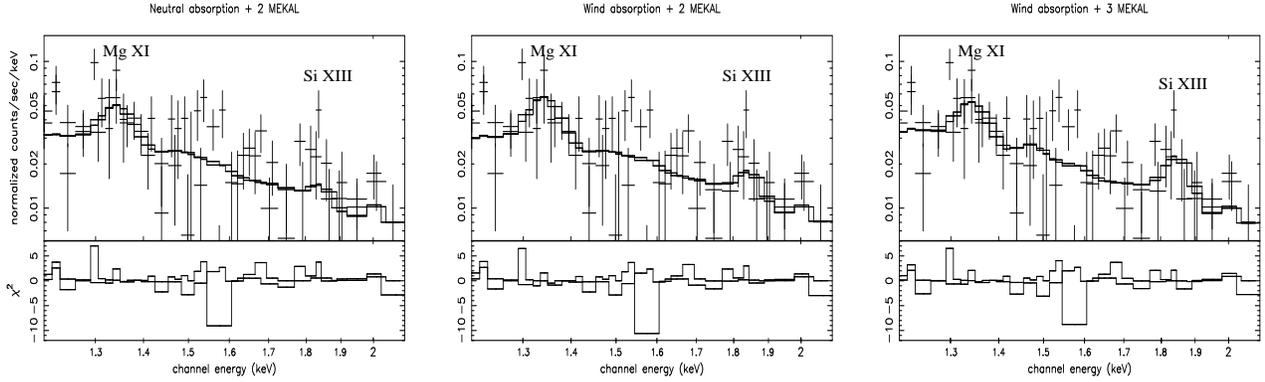}}
\caption{Combined EPIC-MOS1 and EPIC-MOS2 spectra of HD\,168112 fitted with a purely thermal model displayed between 1.2 and 2.1 keV. Data are taken from the April observation. The two model-lines appearing in the upper part are due to the fact that the response matrices of the two instruments are different. The two most prominent lines are \ion{Mg}{xi} at about 1.34 keV and \ion{Si}{xiii} at about 1.86 keV. The ISM absorption component is frozen at N$_\mathrm{H}$ = 0.58\,$\times$\,10$^{22}$\,cm$^{-2}$. Left panel: 
{\tt wabs$_\mathrm{ISM}$*wabs*(2-T mekal)}. Middle panel: {\tt wabs$_\mathrm{ISM}$*wind*(2-T mekal)}. Right panel: {\tt wabs$_\mathrm{ISM}$*wind*(3-T mekal)}.\label{lines}}
\end{center}
\end{figure*}

A third {\tt mekal} component was finally added to the model discussed above. This trial yielded temperatures of about 0.30, 0.74 and 3.6 keV for the EPIC-MOS data simultaneously, and of about 0.28, 0.68 and 3.1 keV for the simultaneous fit of the three EPIC data sets. However, this additional component did not improve the $\chi^2_\nu$ and was therefore not considered in the following.\\
To illustrate the effect of the various models on our fittings, we consider how these models fit specific lines appearing in our EPIC spectra. In the case of the combined fit of EPIC-MOS data for the first observation, Fig.\,\ref{lines} shows the 1.2 to 2.1 keV region, where two of the most prominent lines of the spectrum lie. These lines are \ion{Mg}{xi} at about 1.34 keV and \ion{Si}{xiii} at about 1.86 keV. The three panels of Fig.\,\ref{lines} show respectively the results from the 2-T model using the cold local absorption, the 2-T model using the warm local absorption, and the 3-T model using the warm local absorption. If the first model reproduces the \ion{Si}{xiii} line rather poorly, the situation is significantly improved when the warm material model is used, as seen in the second and third panels. Indeed, for the 2-T model, the soft component temperature is higher in the case of the warm absorption. Consequently, the model describes a plasma whose temperature is closer to the temperature of maximum emissivity ($\sim$\,10$^{7}$\,K) of this \ion{Si}{xiii} line, and the same is true for the 3-T model which leads to even higher temperatures. If the Si abundance is allowed to vary (using a 2T-{\tt vmekal} model available within {\sc xspec}), the fit points to a possible overabundance of Si relative to the solar composition (a factor $<$ 2), giving rise to a better fit of the \ion{Si}{xiii} line without the need to increase the plasma temperature. For the \ion{Mg}{xi} line, the model seems to better reproduce the line strength in the case of the 2-T model with warm absorption as well. Allowing the Mg abundance to vary does not significantly improve the quality of the fit. According to the APED database (Smith \& Brickhouse \cite{SB}) and the SPEX line list (Kaastra et al.\,\cite{ka2}), the region between the two lines discussed hereabove harbours a number of other lines (\ion{Mg}{xii} at about 1.47 keV being the strongest one, many other lines being due to \ion{Fe}{xxiii} and \ion{Fe}{xxiv}). Some of these may be responsible for the residuals in Fig.\,\ref{lines}.

As stated at the beginning of this section, models including a power law component were also investigated. The results that we obtained are listed in Table\,\ref{fitwTP} for combined EPIC-MOS data, and for EPIC-pn, and finally for the three EPIC instruments together. In this model, a {\tt mekal} thermal component is used along with a power law component. Parameters get values which are very consistent between different data sets, and from one exposure to the other. The typical value of the temperature and of the photon index are respectively $\sim$\,0.3\,keV and $\sim$\,3 for all instruments. We may note that this photon index value is significantly larger than the value ($\Gamma \sim$\,1.5) predicted for IC X-ray emission from relativistic electrons accelerated by strong shocks (Chen \& White \cite{CW}). The problem of the  local minima encountered in the case of purely thermal models was even more obvious in the case of this model. As it was the case for the 2-T model, the second local minimum gives a temperature for the thermal component of about 0.5 -- 0.6 keV, with the photon index of the power law nearly unchanged. This local minimum was rejected because of the impossibility to obtain a reasonable value of N$_\mathrm{w}$ for the second exposure, and because of a wider dispersion of the normalization values of all emission components.\\ 

We note that the fittings with either purely thermal or thermal + non-thermal models yield essentially equivalent $\chi^2$. However, the likely presence of a moderately strong Fe-K line at about 6.5\,keV is not accounted for by models including power laws. If the detection of this line is real (it appears in 
EPIC-MOS (see Fig.\,\ref{spmos}) and EPIC-pn spectra (see Fig.\,\ref{pnfit})), at least for the first observation) and is confirmed by future observations, its presence argues in favour of a thermal origin of the high energy tail of the EPIC spectra.\\

A word of caution is however necessary here. In fact, the formation of lines mainly occurs through the collisional excitation produced by ions colliding with free electrons. Non-thermal electrons may therefore also contribute to the excitation of several transitions (see e.g. Gabriel et al.\,\cite{gab}; Mewe \cite{mewe2}). Assuming that the iron K\,$\alpha$ line is mainly produced by the {\it 1\,s -- 2\,p} resonance transition of \ion{Fe}{xxv}, we have calculated the collisional excitation rates following the formalism outlined in Mewe (\cite{mewe2}). It turns out that the ratio between the thermal and non-thermal excitation rates is strongly dependent on the low-energy cut-off energy of the power law distribution of the relativistic electrons. For most values of this cut-off energy, the excitation rates are comparable. As a result, the contribution of non-thermal electrons to the line formation will be of the same order as the fraction of these electrons to the total number of electrons. Therefore, if the relativistic electrons make up a small fraction ($\sim$ 1\,\%) of the total number of free electrons, their contribution to the Fe\,K\,$\alpha$ line should be rather marginal. In summary, the fact that the Fe\,K line and the high energy continuum can both be reproduced by a 2--3 keV thermal component argues in favour of a thermal origin for this high energy component in the HD\,168112 EPIC spectrum.

\begin{table*}[htb]
\caption{Best-fit parameter values for a {\tt wabs$_\mathrm{ISM}$*wind*(mekal+power)} model for the combined EPIC-MOS data, EPIC-pn, and the three EPIC spectra together. The ISM absorption component is frozen at 0.58\,$\times$\,10$^{22}$\,cm$^{-2}$. The local absorption column accounts for the ionization of the wind, with N$_\mathrm{W}$ expressed in cm$^{-2}$. The upper and lower parts of the table refer to the first and second observation respectively. For the power law component, the normalization parameter (Norm$_2$) corresponds to the photon flux at 1 keV. The error bars represent the 1-$\sigma$ confidence interval.\label{fitwTP}}
\begin{center}
\begin{tabular}{ccccccc}
\hline
\hline
	& $\log{\mathrm{N}_\mathrm{w}}$ & kT$_1$ & Norm$_1$ & $\Gamma$ & Norm$_2$ 
& $\chi^2_\nu$ \\
	&  & (keV) &  &  &  & d.o.f. \\
\hline
\hline
MOS1 + MOS2 & 21.86 & 0.30 & 4.37\,$\times$\,10$^{-3}$ & 2.58 & 2.12\,$\times$\,10$^{-4}$ & 0.99 \\ 
	& $\pm$ 0.09 & $\pm$ 0.03 & $\pm$ 3.02\,$\times$\,10$^{-3}$ & $\pm$ 0.32 & $\pm$ 
1.35\,$\times$\,10$^{-4}$ & 174 \\
\hline
pn	& 21.80 & 0.28 & 3.56\,$\times$\,10$^{-3}$ & 3.02 & 3.78\,$\times$\,10$^{-4}$ & 1.05 \\ 
	& $\pm$ 0.15 & $\pm$ 0.03 & $\pm$ 3.94\,$\times$\,10$^{-3}$ & $\pm$ 0.30 & $\pm$ 1.82\,$\times$\,10$^{-4}$ & 136 \\
\hline
MOS1 + MOS2 & 21.84 & 0.29 & 4.18\,$\times$\,10$^{-3}$ & 2.82 & 2.85\,$\times$\,10$^{-4}$ & 1.01 \\ 
+ pn & $\pm$ 0.08 & $\pm$ 0.02 & $\pm$ 2.55\,$\times$\,10$^{-3}$ & $\pm$ 0.22 & $\pm$ 1.15\,$\times$\,10$^{-4}$ & 315 \\
\hline
\hline
MOS1 + MOS2 & 21.71 & 0.26 & 2.57\,$\times$\,10$^{-3}$ & 2.97 & 2.47\,$\times$\,10$^{-4}$ & 0.82 \\ 
	& $\pm$ 0.21 & $\pm$ 0.03 & $\pm$ 3.63\,$\times$\,10$^{-3}$ & $\pm$ 0.27 & $\pm$ 1.15\,$\times$\,10$^{-4}$ & 116 \\
\hline
pn	& 21.89 & 0.26 & 4.55\,$\times$\,10$^{-3}$ & 2.93 & 2.23\,$\times$\,10$^{-4}$ & 1.00 \\ 
	& $\pm$ 0.87 & $\pm$ 0.02 & $\pm$ 3.36\,$\times$\,10$^{-3}$ & $\pm$ 0.32 & $\pm$ 1.41\,$\times$\,10$^{-4}$ & 89 \\
\hline
MOS1 + MOS2 & 21.86 & 0.26 & 4.30\,$\times$\,10$^{-3}$ & 2.93 & 2.33\,$\times$\,10$^{-4}$ & 0.89 \\
+ pn & $\pm$ 0.70 & $\pm$ 0.01 & $\pm$ 2.46\,$\times$\,10$^{-3}$ & $\pm$ 0.22 & $\pm$ 0.96\,$\times$\,10$^{-4}$ & 210\\
\hline
\hline
\end{tabular}
\end{center}
\end{table*}

\subsubsection{X-ray fluxes\label{xflux}}

\begin{table*}
\caption{X-ray fluxes and luminosities of HD\,168112. The third column gives the absorbed flux, and the fourth one yields the flux corrected for interstellar absorption. Luminosities corrected for interstellar absorption were determined for the two exposures between 0.4 and 10.0 keV and are given in the fifth column. These luminosities were calculated adopting a 2 kpc distance (see Table\,\ref{param}).\label{lum}}
\begin{center}
\begin{tabular}{llccc}
\hline \hline
Observation & Data set & Abs.\,flux & Corr.\,Flux & Corr.\,Luminosity \\
	& & (erg\,cm$^{-2}$\,s$^{-1}$) & (erg\,cm$^{-2}$\,s$^{-1}$) & 
(erg\,s$^{-1}$) \\
\hline
April 2002 & MOS1 + MOS2 & 4.77\,$\times$\,10$^{-13}$ & 1.93\,$\times$\,10$^{-12}$ & 9.15\,$\times$\,10$^{32}$ \\
	   & MOS1 + MOS2 + pn & 4.79\,$\times$\,10$^{-13}$ & 1.98\,$\times$\,10$^{-12}$ & 9.38\,$\times$\,10$^{32}$ \\
Sept. 2002 & MOS1 + MOS2 & 3.59\,$\times$\,10$^{-13}$ & 1.84\,$\times$\,10$^{-12}$ & 8.72\,$\times$\,10$^{32}$ \\
	   & MOS1 + MOS2 + pn & 3.44\,$\times$\,10$^{-13}$ & 1.79\,$\times$\,10$^{-12}$ & 8.48\,$\times$\,10$^{32}$ \\
\hline
\end{tabular}
\end{center}
\end{table*}

Based on the best-fit parameters for the two-temperature {\tt mekal} model, we evaluated the fluxes between 0.4 and 10.0 keV. These are listed in Table\,\ref{fitwTT}. The fluxes for the combined EPIC-MOS and EPIC-MOS/pn fit, as well as the corresponding luminosities calculated adopting a 2 kpc distance (Bieging et al.\,\cite{BAC}), are also given in Table\,\ref{lum}. The absorbed X-ray flux suffered a decrease of about 25--30\,\% from April to September. To check for possible systematic errors due to the somewhat different location of the source on the detector from one pointing to the other, we evaluated the exposure map at the position of the source on the detector. Between the two observations, the net exposure difference amounts to about 0.5\,\% for EPIC-MOS1, 2\,\% for EPIC-MOS2, and 3\,\% for EPIC-pn. These differences are too small to explain the amplitude of the flux variation observed for HD\,168112.  We may note that the fluxes obtained in the case of cold material local absorption are similar to those quoted in Table\,\ref{fitwTT}.\\

We compared the luminosities from the two observations to the value expected for a single O star following the relation given by Bergh\"ofer et al.\,(\cite{BSDC}). To do so, we used the bolometric luminosity given by Leitherer (\cite{Leitherer88}), i.e.\ 3.0\,$\times$\,10$^{39}$ erg\,s$^{-1}$, and we obtained an expected L$_\mathrm{X}$ of about 5.3\,$\times$\,10$^{32}$ erg\,s$^{-1}$. As a result, the expected L$_\mathrm{X}$/L$_\mathrm{bol}$ ratio is 1.75\,$\times$\,10$^{-7}$. We then computed the L$_\mathrm{X}$/L$_\mathrm{bol}$ ratio for both exposures using the L$_\mathrm{X}$ values given in Table\,\ref{lum}. These calculations yield 3.1\,$\times$\,10$^{-7}$ and 2.8\,$\times$\,10$^{-7}$ respectively for the two observations. As a consequence, we detect no significant luminosity excess. Indeed, for both observations the X-ray level is within a factor 1.7 of what is expected for a single star. This is well within the intrinsic dispersion of the relation given by Bergh\"ofer et al.\,(\cite{BSDC}), which is about a factor 2.\\ 

\section{Variability of the X-ray flux \label{var}}

\begin{table}
\caption{EPIC-MOS and EPIC-pn count rates obtained for HD\,168112 during the two {\it XMM-Newton} observations between 0.4 and 10.0 keV. The error bars represent the 1-$\sigma$ confidence interval.\label{cr}}
\begin{center}
\begin{tabular}{lccc}
\hline \hline
	& \multicolumn{3}{c}{EPIC} \\
\cline{2-4}
Observation & MOS1 & MOS2 & pn \\
	& (cts\,s$^{-1}$) & (cts\,s$^{-1}$) & (cts\,s$^{-1}$) \\
\hline
April 2002 & 0.060 & 0.057 & 0.157 \\
  & $\pm$ 0.003 & $\pm$ 0.003 & $\pm$ 0.006 \\
Sept. 2002 & 0.037 & 0.039 & 0.111 \\
  & $\pm$ 0.002 & $\pm$ 0.002 & $\pm$ 0.004 \\
\hline
\end{tabular}
\end{center}
\end{table}

As mentioned in Sect.\,\ref{xflux}, HD\,168112 shows a substantial decrease in its observed flux from the first observation to the second one. This reduction in the X-ray emission is clearly shown by the count rates in Table\,\ref{cr} that drop between the two exposures. Considering the mean combined EPIC-MOS count rate, the decrease is of the order of 35\,\% compared to the first observation value.\\

The flux determinations in the previous section require a more detailed discussion. Between 0.4 and 10.0 keV, it appears that the variability observed for the combined EPIC-MOS and EPIC-MOS/pn data is about 25 -- 30\,\% for absorbed fluxes, whilst the dereddened fluxes vary by only about 5 -- 10\,\% (see Table\,\ref{lum}). However, at energies above $\sim 0.8$\,keV, the absorbed and dereddened fluxes display the same level of variability. The discrepancy in the variability of flux over the entire energy range arises therefore from the lowest energy bins. In fact, the correction for the interstellar absorption amounts to a factor about 35 to 100 at energies of 0.5 -- 0.7\,keV. Thus, although the energy bins below 0.8\,keV contribute only to a marginal fraction of the total observed flux (of the order 5\%), they account for almost half of the total dereddened flux when corrected for the ISM absorption. Therefore, any variations occurring predominantly at higher energy (typically above 0.8 keV), will be diluted while considering the whole EPIC bandpass of an ISM absorption corrected spectrum and will consequently lose most of its significance. 
Moreover, fluxes estimated in the soft part of the spectrum are very sensitive to the quality of the data, which in our case is quite poor because of the low count rate in this energy band. We may also emphasize that a moderate uncertainty on the ISM column density could have a considerable impact on the dereddened flux in the lowest energy bins. Considering all these limitations, we decided to focus our discussion on the energy range above 0.8\,keV that is less drastically affected by the correction of the ISM absorption.

Table\,\ref{fluxes} gives the observed (absorbed) fluxes estimated from the two temperature model with the warm wind absorption fitted to the combined EPIC-MOS data. The quoted fluxes are given in three energy bands that were chosen to clearly illustrate the dependence of the variability as a function of the energy band. The softer band (0.8 -- 1.0 keV) displays a variability of about 20\,\%, and between 1.0 and 5.0 keV the variability is about 28\,\%\footnote{We do not consider flux values above 5.0 keV for two reasons: firstly only a few photons are detected in that region of the spectrum, and secondly this energy band is the most strongly affected by the soft proton flare which occurred during the April observation.}. These values suggest that the variability is somewhat stronger in the hard part of the spectrum. We may also note that the reported variability levels remain unchanged within one or two percent, if we consider fluxes corrected for the interstellar medium.\\

\begin{table}
\caption{Observed fluxes estimated in the case of the model resulting from combined EPIC-MOS data fitted with the two-temperature model, in the case of the warm local absorption model. \label{fluxes}}
\begin{center}
\begin{tabular}{cccc}
\hline \hline
Energy & April 2002 & Sept. 2002 & Variability\\
(keV) & (erg\,cm$^{-2}$\,s$^{-1}$) & (erg\,cm$^{-2}$\,s$^{-1}$) &  \\
\hline
0.8 -- 10.0 & 4.61\,$\times$\,10$^{-13}$ & 3.30\,$\times$\,10$^{-13}$ & $\sim$\,27\,\% \\
0.8 -- 1.0 & 0.66\,$\times$\,10$^{-13}$ & 0.53\,$\times$\,10$^{-13}$ & $\sim$\,20\,\% \\
1.0 -- 5.0 & 3.24\,$\times$\,10$^{-13}$ & 2.33\,$\times$\,10$^{-13}$ & $\sim$\,28\,\% \\
\hline
\end{tabular}
\end{center}
\end{table}

\begin{figure*}[htb]
\begin{center}
\resizebox{16.0cm}{11cm}{\includegraphics{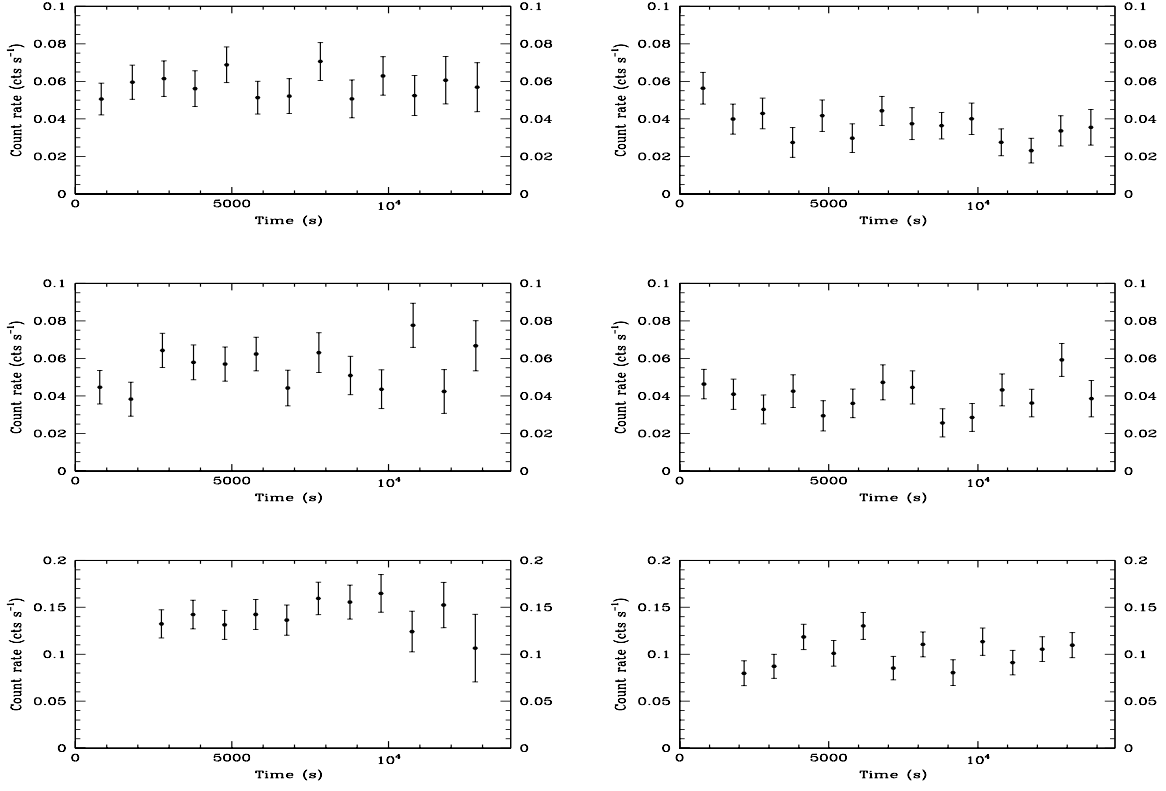}}
\caption{Light curves obtained for a 1000 s time bin between 0.4 and 10.0 keV for observation 1 (on the left) and observation 2 (on the right). No correlation is found between the fluctuations observed for the three EPIC instruments. Top panel: MOS1. Middle panel: MOS2. Bottom panel: pn. The error bars represent the 1-$\sigma$ poissonian standard deviation.\label{lc}}
\end{center}
\end{figure*}

Beside the variability observed between the two {\it XMM-Newton} observing periods, we also looked for short term variations (i.e. variations within a single exposure). We extracted light curves in the four energy bands, respectively (1) 0.4 -- 1.0 keV, (2) 1.0 -- 2.5 keV, (3) 2.5 -- 10.0 keV and 0.4 -- 10.0 keV. For each energy range, event lists were binned with 200\,s, 500\,s, 1000\,s, and 2000\,s time intervals. For each time bin of the light curves, and for each energy range (expressed in PI channel numbers, considering that 1 PI channel corresponds approximately to 1 eV), count rates were calculated with their corresponding standard deviation. These light curves were extracted within the same source and background regions used for spectral analysis (Sect.\,\ref{sect_epic1}), and all source light curves were corrected for background and Good Time Intervals (GTIs)\footnote{Even if no flare contaminated time interval was rejected, standard GTIs are anyway always applied to {\it XMM-Newton} data, and must be taken into account in every timing analysis.}. Fig.\,\ref{lc} shows light curves obtained for a time bin of 1000 s between 0.4 and 10.0 keV for EPIC-MOS and EPIC-pn instruments for the two observations. We do not see any correlation between the fluctuations of the count rate of the three instruments. Moreover, variability tests ($\chi^2$, Kolmogorov-Smirnov, and pov-test as described by Sana et al.\,\cite{sana}) applied to all our data sets did not reveal any significant variability on time scales shorter than the duration of an exposure, except the one due to the soft proton flare of the first observation (see Sect.\,\ref{sect_flare}).\\

\section{Comparison with previous X-ray observations \label{former}}
Prior to the present {\it XMM-Newton} observations, HD\,168112 was observed with {\it EINSTEIN} (Chlebowski et al.\,\cite{Chle}) and the estimated X-ray luminosity (between 0.2 and 3.5\,keV) was about 2.04\,$\times$\,10$^{33}$\,erg\,s$^{-1}$ for a 2\,kpc distance. The {\it EINSTEIN} count rate was about 0.018 $\pm$ 0.004\,cts\,s$^{-1}$, for a 7591.5\,s exposure ({\it EINSTEIN} sequence number 5960; observed on March 14, 1981).

Because of the differences in the adopted model hypotheses for the {\it EINSTEIN} and our luminosity estimates, X-ray luminosities are not directly comparable. To achieve a more suitable comparison, we folded our {\it XMM-Newton} best fit 2-T models through the response matrix of the IPC instrument. The corresponding equivalent IPC count rates estimated between 0.2 and 3.5 keV are 0.013 $\pm$ 0.003\,cts\,s$^{-1}$ and 0.009 $\pm$ 0.003\,cts\,s$^{-1}$ respectively for our two exposures. The results for both exposures are thus lower than the actual IPC count rate given by Chlebowski et al.\ (\cite{Chle}), indicating that the flux during the {\it EINSTEIN} observation was higher than at the time of our two {\it XMM-Newton} pointings.

We also retrieved {\it ROSAT}-PSPC data (rp500298n00, 9.5\,ks; obtained 
between September 13 and September 15, 1993) and we inferred a count rate of 0.054 $\pm$ 0.003\,cts\,s$^{-1}$. The same procedure as described hereabove for {\it EINSTEIN} data but using the {\it ROSAT}-PSPC response matrix was applied. The equivalent PSPC count rates corresponding to our two {\it XMM-Newton} observations are respectively 0.027 $\pm$ 0.002\,cts\,s$^{-1}$ and 0.019 $\pm$ 0.002\,cts\,s$^{-1}$. These results suggest that the {\it ROSAT}-PSPC observation of HD\,168112 was performed at a time when the star was in a higher emission state than during the {\it EINSTEIN} and {\it XMM-Newton} observations.

Finally, according to the ROSHRI database (see e.g.\ ledas-www.star.le.ac.uk/rosat/rra), the vignetting corrected count rate of a {\it ROSAT}-HRI observation performed between September 12 and October 9, 1995 (rh201995n00, 36.3 ks) is (7.8 $\pm$ 0.6)\,$\times$\,10$^{-3}$\,cts\,s$^{-1}$. The same procedure as for PSPC was applied using the HRI response matrix and we obtain equivalent HRI count rates of (10.4 $\pm$ 0.1)\,$\times$\,10$^{-3}$ and (7.6 $\pm$ 0.1)\,$\times$\,10$^{-3}$\,cts\,s$^{-1}$ respectively for our two {\it XMM} exposures, suggesting that during the {\it ROSAT}-HRI observation HD\,168112 was in an X-ray emission state similar to what was observed with {\it XMM-Newton} in September 2002.\\

\begin{figure}[ht]
\begin{center}
\resizebox{8.5cm}{4.0cm}{\includegraphics{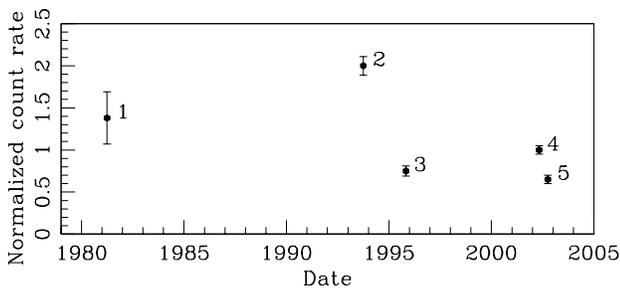}}
\caption{Normalized equivalent X-ray count rates arising from different observatories, as a function of time. 1: {\it EINSTEIN}-IPC, March 1981. 2: {\it ROSAT}-PSPC, September 1993. 3: {\it ROSAT}-HRI, September and October 1995. 4: {\it XMM-Newton}-EPIC, April 2002. 5: {\it XMM-Newton}-EPIC, September 2002.\label{arch}}
\end{center}
\end{figure}

In summary, as a result of this comparison with previous observations, we see that the X-ray flux of HD\,168112 undergoes a quite strong long-term variability. Figure\,\ref{arch} shows the X-ray count rate from these 
observations, normalized relative to the {\it XMM-Newton} count rate in April 2002 arbitrarily set to unity. It is clear from Fig.\,\ref{arch} that the data coverage is insufficient to provide a detailed description of this behaviour. X-ray monitoring of HD\,168112 at a sampling time scale of a few weeks or months would certainly help to better constrain the phenomenon.\\  

\section{VLA observations \label{sect_radio}}

For both epochs, HD\,168112 is detected at 3.6 and 6\,cm (see Fig.~\ref{figure 
VLA maps}), but not at 18 and 20\,cm. For the detections, we measured the flux by fitting an elliptical Gaussian to the source. The Gaussian has the same shape as the cleaned beam, so the position and total intensity are the only parameters fitted. The error bar on the flux due to {\em random} errors is given by the 
root-mean-square (RMS) in the map. To get a feeling for the {\em systematic} errors, we repeated the reduction, systematically dropping one antenna, using different weightings of distant visibilities, or natural weighting instead of robust uniform, or doubling or halving the number of clean components. The error bars given in Table~\ref{table VLA flux results} show the range of results found by these various reductions, or the RMS error bar when it is larger. For each observation, we also compared the peak intensity of the source to the flux derived from the Gaussian fit. There is very good agreement for March data, showing that the sources are indeed point sources. For September, however, the peak intensities are higher than the Gaussian fluxes by about 1-$\sigma$. The error bars of the 3.6 and 6\,cm observations include a 2~\% calibration error (Perley \& Taylor \cite{Perley+Taylor99}).

\begin{figure}
\begin{center}
\includegraphics[scale=0.5]{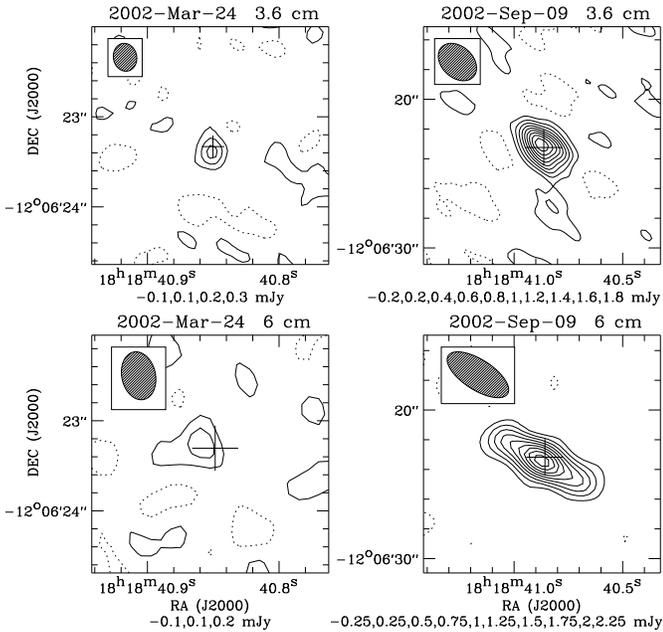}
\caption{VLA maps of HD\,168112 at 3.6 and 6\,cm for both epochs. Note the different spatial scales used.
The synthesized beam is shown in the upper left corner of each panel.The cross indicates the optical ICRS 2000.0 position (from SIMBAD).Contour levels are listed at the bottom of each panel. The first contour line is at a flux levelof $\sim 2 \times$ the RMS of the map. The negative contour is indicated by the dashed line.}
\label{figure VLA maps}
\end{center}
\end{figure}

For the non-detections at 18 and 20 cm, Table~\ref{table VLA flux results} gives an upper limit of three times the RMS measured in a box of $100 \times 100$ pixels centered on HD\,168112. Because of interference at 18\,cm, the RMS at that wavelength is higher than at 20\,cm. The influence of the background structure on the 18\,cm and 20\,cm maps for September 2002 is rather limited. Various combinations of dropping the shortest baselines and depth of cleaning give maps that are acceptably smooth, showing that all background effects have been removed. The upper limits listed in Table~\ref{table VLA flux results} are based on the highest RMS found in the set of acceptable maps.

The major new result from the present observations is the clear difference in flux level between both epochs. The fluxes of September are roughly in agreement with some of the Bieging et al.\ (\cite{BAC}) observations. The March fluxes, however, are substantially lower than their observations by a factor of 5--7.

\begin{table*}
\caption{VLA fluxes and spectral index.}
\label{table VLA flux results}
\begin{tabular}{lllllllrl}
\hline \hline
& \multicolumn{1}{c}{Date} & \multicolumn{5}{c}{Flux (mJy)} & 
   \multicolumn{1}{c}{Spectral} & \multicolumn{1}{c}{Reference} \\
\cline{3-7}
& & \multicolumn{1}{c}{2\,cm} & \multicolumn{1}{c}{3.6\,cm} & 
    \multicolumn{1}{c}{6\,cm} & \multicolumn{1}{c}{18\,cm} & 
     \multicolumn{1}{c}{20\,cm} & \multicolumn{1}{c}{index} \\
\hline
\multicolumn{2}{c}{{\em data from this paper}}\\
\hline
& 2002-Mar-24 & \multicolumn{1}{c}{---} & $0.3 \pm 0.1$ & $0.3 \pm 0.1$ & 
   \multicolumn{1}{c}{---}  & $<0.6$ & $ 0.0 \pm 0.9$\\
& 2002-Sep-11 & \multicolumn{1}{c}{---} & $1.8 \pm 0.2$ & $2.1 \pm 0.3$ & 
   $<2.4$ & $<1.2$ & $-0.3 \pm 0.4$ \\
\hline
\multicolumn{2}{c}{{\em data from literature}}\\
\hline
& 1984-Mar-09 & \multicolumn{1}{c}{---} & \multicolumn{1}{c}{---} & $1.3 \pm 0.1$ 
& 
   \multicolumn{1}{c}{---} & \multicolumn{1}{c}{---} & \multicolumn{1}{c}{---} &
   Bieging et al.\ (\cite{BAC}) \\
& 1984-Apr-04 & $1.2 \pm 0.2$ & \multicolumn{1}{c}{---} & $1.9 \pm 0.1$ & 
   \multicolumn{1}{c}{---} & \multicolumn{1}{c}{---} & $-0.4 \pm 0.2$ & 
   Bieging et al.\ (\cite{BAC}) \\
& 1984-Dec-21 & $1.3 \pm 0.1$ & \multicolumn{1}{c}{---} & $5.4 \pm 0.1$ & 
   \multicolumn{1}{c}{---} & $8.2 \pm 0.3$ & $-0.8 \pm 0.5$ & 
   Bieging et al.\ (\cite{BAC}) \\
& 1989-Nov-10 & \multicolumn{1}{c}{---} & \multicolumn{1}{c}{---} & 
   \multicolumn{1}{c}{---} & $<2.4$ & \multicolumn{1}{c}{---} & 
\multicolumn{1}{c}{---} 
   & Phillips \& Titus~(\cite{Phillips+Titus90}) \\
\hline
\end{tabular}
\end{table*}

We explored the possibility that these low fluxes are due to poor phase calibration (Thompson et al.\,\cite{Thompson+al}, pp. 428--432). Phase errors increase with baseline length, so the A configuration used in March is the most sensitive to this effect. At 3.6\,cm, most gain phases between the two calibration observations bracketing HD\,168112 change by $< 20$\degr, with only a few showing higher values up to 50\degr. At 6\,cm, the gain phases change by about 30\degr. Comparison with previous work (Blomme et al.\,\cite{Blomme+al02}, \cite{Blomme+al03}) suggests that, for such changes, some flux might be lost, but not a substantial amount. The effect is certainly not large enough to explain the factor 5--7 difference in the fluxes.

Another possibility is that we could be resolving the star. However, Blomme et al.\ (\cite{Blomme+al03}) showed that for the much closer star $\zeta$~Pup the flux lost in this way only amounts to $\sim$~10\%. Furthermore, one of a number of unpublished VLA archive data taken in C configuration shows a flux for HD\,168112 that is as low as our March observation: $0.3 \pm 0.05$\,mJy at 3.6\,cm (Blomme et al.\,2004, in preparation). One may also wonder whether the change in resolution between the two observations could explain the flux variation. As mentioned above, low fluxes were already observed with the C configuration, although our high flux measurement was also obtained with the C configuration. Moreover, by throwing away some of the data, a lower-resolution observation can be simulated with the March data, but that does not change the flux significantly.\\

Table~\ref{table VLA flux results} also lists the spectral index ($\alpha$, where $F_\nu \propto \lambda^{-\alpha}$) based on our 3.6 and 6\,cm observations. For September, $\alpha$ is significantly 
different from the value expected for thermal emission (+0.6), showing that the emission at that time was definitely non-thermal. For March, the result is less clear due to the large error bar on $\alpha$. It is possible that at that time HD\,168112 was showing a thermal spectrum. However, using the mass loss rate, terminal velocity and distance from Table\,\ref{param}, the Wright \& Barlow~(\cite{Wright+Barlow75}) formula predicts a 6\,cm flux of only 0.03\,mJy. Even the observed flux in March is a factor ten larger than this value. Therefore, we are definitely not seeing the underlying free-free emission of the stellar wind.

Combining our data with published values shows considerable variability ($0.3-5.4$\,mJy at 6\,cm and  $<1.2 - 8.2$\,mJy at 20\,cm). This variability in fluxes and in spectral index is consistent with the way other non-thermal emitters, such as Cyg\,OB2 No.\,9 (Waldron et al.\,\cite{Waldron+al98}) and WR\,140 (Williams et al.\,\cite{Wil2}), behave.

\section{Optical observations \label{sect_opt}}
\begin{figure*}
\resizebox{\hsize}{!}{\includegraphics{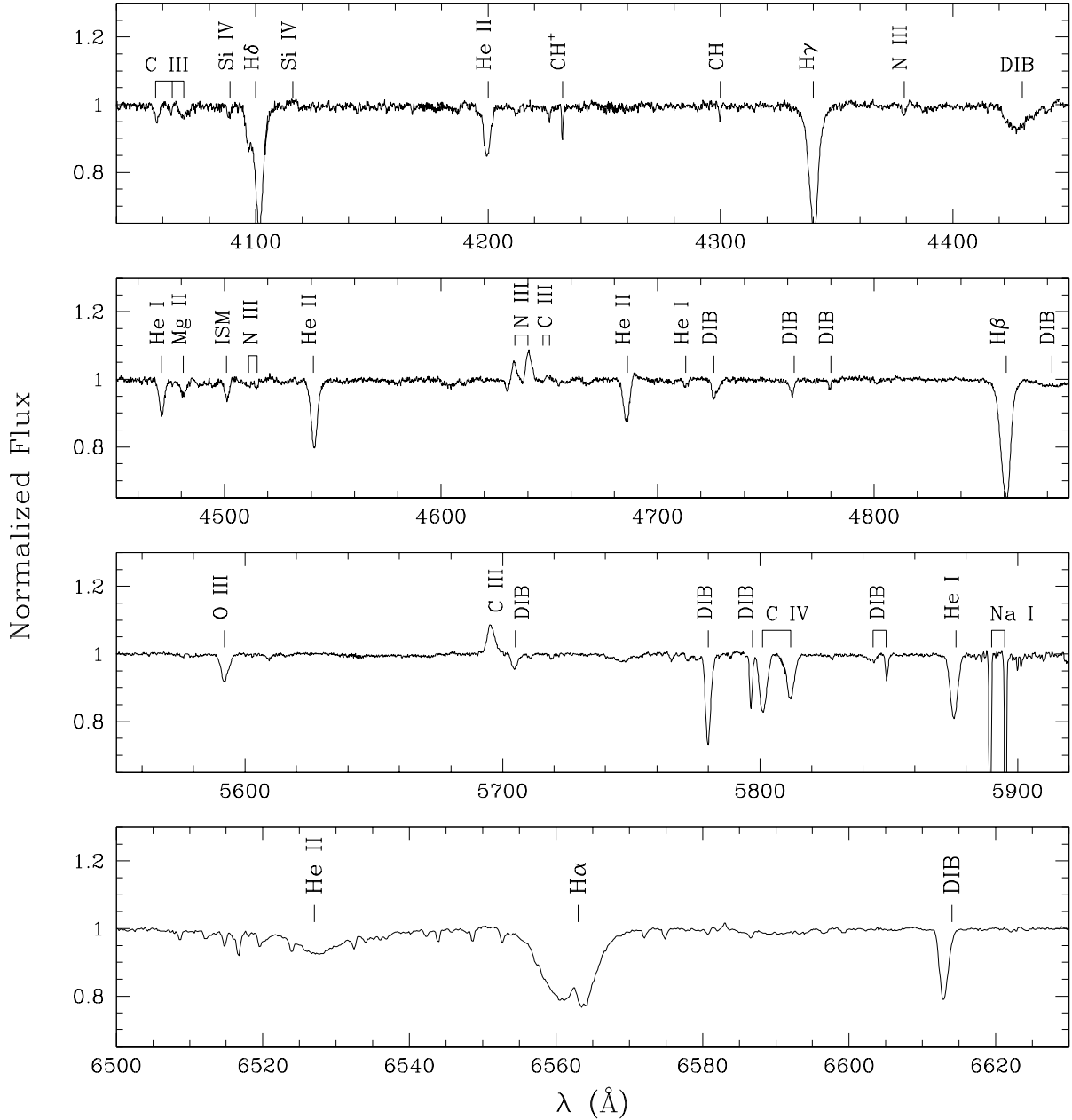}}
\caption{Mean optical spectrum of HD\,168112 as observed from SPM in April 2003. The most prominent stellar and interstellar lines as well as diffuse interstellar bands (DIBs) are indicated. \label{fig-spec}}
\end{figure*}

The mean spectrum of HD\,168112 as observed at SPM is shown in Fig.\,\ref{fig-spec}. The most prominent absorption lines belong to H\,{\sc i}, He\,{\sc i}, He\,{\sc ii} as well as to some metal ions (e.g.\ O\,{\sc iii}, C\,{\sc iii}, C\,{\sc iv}, N\,{\sc iii},...). The equivalent width (EW) ratio of the He\,{\sc i} $\lambda$\,4471 and He\,{\sc ii} $\lambda$\,4542 absorption lines ($W' = {\rm EW}(4471)/{\rm EW}(4542) = 0.41$) yields an O5.5 spectral type according to the criterion of Conti (\cite{conti73}). The strength of the 
He\,{\sc ii} $\lambda$\,4686 absorption (${\rm EW} = 0.49$\,\AA) as well as its slightly asymmetric shape with a steeper red wing suggest a giant luminosity class.

There are also a few narrow emission lines: N\,{\sc iii} $\lambda\lambda$\,4634--41 with equivalent widths of $-0.17$ and $-0.26$\,\AA\ respectively, as well as C\,{\sc iii} $\lambda$\,5696 (EW = $-0.30$\,\AA). The Si\,{\sc iv} $\lambda$\,4116 line is also seen as an extremely faint emission (EW $\sim -0.08$\,\AA), whereas Si\,{\sc iv} $\lambda$\,4089 is seen as a very faint absorption (EW $\sim 0.07$\,\AA). This situation is similar to what is found in 9\,Sgr (Rauw et al.\,\cite{9sgr}) and is usually interpreted as the $\lambda$\,4089 absorption and emission almost canceling each other (see Walborn \cite{Wal2}). We note also the presence of the two very weak S\,{\sc iv} emissions at 4485 and 4504\,\AA\ (Werner \& Rauch \cite{WR}) and of a weak C\,{\sc iii} $\lambda\lambda$ 4647--50 blend. Finally, we note that the core of the H$\alpha$ absorption line is partially filled in by emission. The EW of the H$\alpha$ line as measured on our spectra ($1.94 \pm 0.11$\,\AA) is in excellent agreement with the value (2\,\AA) adopted by Leitherer (\cite{Leitherer88}).

Considering these spectral properties, we classify HD\,168112 as O5.5\,III(f$^+$), in excellent agreement with classifications published in the literature (see \ Mathys \cite{mathys} and references therein).\\

\begin{table}[htb]
\caption[]{\label{tab: diary} Average radial velocities of nine absorption lines in the spectrum of HD\,168112 as measured on our SPM and FEROS data. The third column yields the instrumentation used for the observation: echelle spectrograph at San Pedro (SPM), FEROS at the 1.5\,m telescope (F1.5) or FEROS at the 2.2\,m 
telescope (F2.2).}
\begin{center}
\begin{tabular}{c c c}
\hline \hline
\vspace*{-3mm}\\
\multicolumn{1}{c}{Date} & $\overline{RV}$ & Inst.\\
\multicolumn{1}{c}{HJD $-$ 2450000} & (km\,s$^{-1}$) & \\
\hline
2040.746 & 15.8 & F1.5 \\
2382.900 & 21.1 & F1.5 \\
2745.919 & 12.1 & SPM \\
2746.875 & 18.4 & SPM \\
2747.867 & 11.3 & SPM \\
2748.842 & 20.8 & SPM \\
2749.876 &  7.2 & SPM \\
2750.864 & 13.2 & SPM \\
2783.820 & 16.4 & F2.2 \\
2784.815 & 18.4 & F2.2 \\
\hline
\end{tabular}
\end{center}
\end{table}

\begin{table*}
\caption[]{\label{tab-ltpv} Differential LTPV photometry of HD\,168112. The quoted magnitudes and 1-$\sigma$ dispersions correspond to the difference A\,5020 $-$ B\,5020 (where A\,5020 is the magnitude of HD\,168112 and B\,5020 that of HD\,168135). The first column yields the code of the telescope + instrument configuration (see Manfroid et al.\,\cite{manfroid}) and the second column indicates the number of data points for each set-up.}
\begin{center}
\begin{tabular}{c c c c c c}
\hline \hline
\vspace*{-3mm}\\
\multicolumn{1}{c}{Telescope} & N & $u$ & $v$ & $b$ & $y$\\
\hline
5 & 26 & $0.780 \pm 0.011$ & $1.205 \pm 0.005$ & $0.976 \pm 0.004$ & $0.570 \pm 
0.005$ \\
6 & 39 & $0.747 \pm 0.053$ & $1.192 \pm 0.012$ & $0.962 \pm 0.008$ & $0.568 \pm 
0.010$ \\
7 & 30 & $0.873 \pm 0.010$ & $1.252 \pm 0.010$ & $0.966 \pm 0.010$ & $0.580 \pm 
0.009$ \\
\hline
\end{tabular}
\end{center}
\end{table*}

So far, there exists basically no information on the multiplicity of HD\,168112 in the literature. We have measured the heliocentric radial velocities (RVs) of the strongest absorption lines in our spectra. We then averaged the RVs of nine of them (H$\gamma$, H$\beta$, He\,{\sc ii} $\lambda\lambda$\,4200, 4542, 4686, 
5412; He\,{\sc i} $\lambda$\,5876 and C\,{\sc iv} $\lambda\lambda$\,5801, 5812). The rest wavelengths of these lines were adopted from Underhill (\cite{underhill}).

The results listed in Table\,\ref{tab: diary} do not reveal any significant RV changes. For instance, the mean RV averaged over the SPM data set amounts to $13.8 \pm 5.0$\,km\,s$^{-1}$, whilst this mean is $15.5 \pm 4.5$\,km\,s$^{-1}$ over the entire data set. No obvious trend appears in the RVs over the six nights of the SPM campaign or between the various FEROS campaigns. The lack of significant RV variations suggests that HD\,168112 is either single or it could be a binary system with a rather large mass ratio seen under a low inclination angle. 

Finally, we also consider archive photometric data of HD\,168112. A set of such data was collected between 1982 and 1990 in the framework of the Long Term Photometry of Variables (LTPV) programme at ESO (Manfroid et 
al.\,\cite{manfroid}, Sterken et al.\,\cite{sterken}). HD\,168112 was used as a comparison star designated A\,5020 in the context of the LTPV. The data were obtained in the Str\"omgren system with various telescopes and photometers at the La Silla observatory (see Manfroid et al.\,\cite{manfroid}). We have retrieved the data from this campaign and we analysed the differential photometry of A\,5020 $-$ B\,5020, where B\,5020 is a second comparison star (= HD\,168135, B9\,V) that was observed immediately after or before HD\,168112. Since each telescope and instrumental configuration has its specific properties, we analysed the corresponding data separately. The results are presented in Table\,\ref{tab-ltpv}. The 1-$\sigma$ dispersions of the differential magnitudes are always small (except perhaps for the $u$ filter and instrumental system 6) and are comparable to typical results of the LTPV campaign (Manfroid et al.\,\cite{manfroid}). There is no significant trend in the data points either. We conclude that the existing photometric data do not display any significant 
variations that could be attributed e.g.\ to the effects of binarity.\\

In summary, our optical analysis failed to reveal any evidence of binarity on time scales of weeks, or from one year to the next. Moreover, photometric data do not show any variation either. These results suggest three possible scenarios. Firstly, the binary system, if any, could be seen under a very low inclination angle, preventing us from seeing any radial velocity or photometric variations. Secondly, the orbital period of the suspected binary system could be very long, or the sampling of our RV measurements could be too sparse to reveal the actual RV change. Thirdly, HD\,168112 could be a single star.\\

\section{Discussion \label{disc}}
  
\subsection{Scenarios for a thermal hard X-ray emission from HD\,168112 \label{ther}}

It appears that the bulk of the X-rays have a thermal origin. Indeed the continuum is well fitted by bremsstrahlung emission, while there is clear evidence for thermal line emission (e.g. \ion{Mg}{xi} at 1.34 keV, \ion{Si}{xiii} at 1.86 keV and the Fe\,K blend at 6.5 keV). While a non-thermal component may also make a small contribution (this possibility is discussed further in Sect.\,\ref{nts}), we will concentrate first on possible explanations for the observed thermal emission.\\

\subsubsection{The single star scenario} 

Two features of the X-ray data of HD\,168112 are rather unusual for a single O-star. On the one hand, the X-ray flux appears to be variable by a factor of a few. Such a variability is not expected for presumably single O-stars, as stated by Bergh\"ofer \& Schmitt (\cite{BS}) on the basis of {\it ROSAT} data. On the other hand, all our spectral fittings with two temperature thermal models suggest plasma temperatures for the hard component of the order of 2--3 keV, i.e. in the range 2.3--3.5\,10$^{7}$\,K. The X-ray emission for single hot stars is usually attributed to the effects of instabilities and shocks arising in their stellar winds. While these shocks may produce some emission at energies comparable to those seen in HD\,168112, the emission measure associated with this very hot plasma is likely to be rather low. More importantly, when we compare our EPIC spectra with X-ray spectra of presumably single O-stars such as $\zeta$\,Pup (Kahn et al.\,\cite{Kahn}), we clearly notice that the latter are dominated by a soft thermal emission. In order to check whether this impression could be biased by the much lower interstellar column towards $\zeta$\,Pup (compared to HD\,168112), we have extracted the {\it ASCA} SIS spectra of this star\footnote{$\zeta$\,Pup was observed with the {\it ASCA} satellite in October 1993 during about 30\,ks (P.I.: Dr. T.G. Tsuru). Screened event files were retrieved from the ARNIE database at Leicester University and reduced using the {\sc xselect} software (version 1.4b).}. The spectrum was fitted with a thermal model with a temperature of about 0.65 keV, and this model was then reddened by the same ISM column as HD\,168112. Clearly this fake reddened $\zeta$\,Pup spectrum does not exhibit a hard tail similar to that seen in HD\,168112. We therefore conclude that such a hard emission is not seen in the spectrum of a presumably single star like $\zeta$\,Pup.\\

One way to account for the high plasma temperatures derived in our fittings could possibly result from the confinement of the wind by a magnetic field. Babel \& Montmerle (\cite{BM}) suggest that a moderate magnetic field could deviate the stellar winds from either hemisphere towards the magnetic equator, where they would collide, thereby producing a plasma at temperatures as high as several $10^7$\,K. In the case of HD\,168112, a magnetic field is indeed expected to be present since we observe synchrotron radio emission.\\

\subsubsection{The binary scenario}

A striking property of the X-ray emission of HD\,168112 is its strong 
variability between the two observations. The X-ray flux suffered a decrease of about 25-30 \% between April 2002 and September 2002. Comparison with archive data confirms the existence of a substantial variability of the X-ray flux of HD\,168112. This long term variability is compatible with a binary scenario for HD\,168112. This idea is reinforced by the fact that this variability is mostly observed in the hard EPIC bandpass, i.e. in the part of the spectrum presumably mostly influenced by a putative wind-wind collision X-ray emission in a O+O-type binary. In a colliding wind scenario, we do not expect strong variability in the soft part of the spectrum as it is mostly produced by the intrinsic emission from both components. In this context, the observed variability could be due for instance to a significant orbital eccentricity. In fact, the emission measure of the hot post-shock plasma in an adiabatic wind collision zone varies roughly as $1/D$ where $D$ is the orbital separation of the binary components (Stevens et al.\ \cite{SBP}; Pittard \& Stevens\ \cite{PS}). Therefore, an orbital eccentricity of e\,$\geq$\,0.5 would allow to account for the factor 3 variability of the X-ray fluxes as seen in Fig.\,\ref{arch}. The variations seen in the latter figure are compatible with a possible orbital period of several years. Moreover, the fact that this variability seems to be most significant at higher energies is compatible with a variation of an X-ray emission component originating from a wind-wind interaction.\\

Finally, we emphasize that the temperatures of the hard component in the 2-T fittings imply pre-shock velocities of the order of 1300 -- 1600 km\,s$^{-1}$. Such values are indeed typical for relatively wide colliding wind binaries (Pittard \& Stevens\,\cite{PS}) where the winds collide with velocities near v$_{\infty}$.

\subsection{HD\,168112 as a non-thermal emitter\label{nts}}

HD\,168112 is known to be a variable non-thermal radio emitter, and this is reconfirmed by our new data. Combining our results with those of Bieging et al.\ (\cite{BAC}), we find spectral indices that vary between 0.0 and $-0.8$ with a 6\,cm flux ranging between 0.3\,mJy and 5.4\,mJy. As non-thermal processes require the existence of relativistic electrons, thermal electrons must be accelerated to relativistic velocities. This could in principle be done by the first order Fermi mechanism: it requires an acceleration time of only a few minutes to generate relativistic electrons that emit synchrotron radiation in the observable radio region (Van Loo et al.\,\cite{vanloo}). Both synchrotron emission and Fermi mechanism require the existence of a moderate magnetic field (e.g.\ Longair \cite{Long}). Magnetic fields associated with OB stars have been measured using spectropolarimetric methods in the case of $\beta$\,Cep and $\theta^1$\,Ori\,C (Donati et al.\,\cite{Don1}, \cite{Don2}). The acceleration of relativistic electrons by the Fermi mechanism can occur either in the shocks due to intrinsic instabilities of the wind of a single star or in a colliding wind binary, and in the following sections we discuss these models in greater depth.\\

\subsubsection{The single star scenario}
At the present stage of the study of HD\,168112, this star is not known as a binary system. As stated in the introduction, the only known mechanism for a single O-star to accelerate relativistic electrons is by means of hydrodynamic shocks inside its stellar wind. However, since the optical depth of the wind at radio wavelengths is very large, the radio photosphere is located far from the visible photosphere and relativistic electrons accelerated in the inner regions of the wind would have to travel over large distances before they could produce observable synchrotron radio emission. Due to the intense UV stellar radiation field, the time scale for IC thermalization of relativistic electrons is quite short. This implies that the relativistic electrons responsible for the observed synchrotron emission must be accelerated in situ, i.e.\ the shocks responsible for this acceleration must subsist at large distances from the stellar surface (Chen \& White \cite{CW2}). On the other hand, shocks occuring in the inner wind where the UV flux is strongest could also accelerate electrons to relativistic energies. These relativistic electrons could produce a power law tail in the hard X-ray to soft $\gamma$-ray energy range through IC scattering (Chen \& White \cite{CW}, \cite{CW1b}). However, one should emphasize that in this case the population of relativistic electrons responsible for the non-thermal X-rays would be distinct from that producing the synchrotron emission.

Our EPIC spectral fittings with power law models yield photon indices of the order of 3. In the framework of the first order Fermi mechanism, the index of the electron distribution is related to the shock properties through the following relation: $n = (\chi+2)/(\chi-1)$ where $n$ is the index of the relativistic electrons population and $\chi$ is the compression ratio, defined as the ratio between the velocity of the preshock and postshock regions (Bell\,\cite{Bella},\,\cite{Bellb}). This electron index is related to the photon index of the IC emission as follows: $\Gamma = (n+1)/2$, where $\Gamma$ is the photon index. For a photon index of $\approx$ 3, this leads to $n \approx$ 5 and $\chi \approx 1.75$. We note that the value for strong shocks is $\chi$\,=\,4. If the hard X-rays in the spectrum of HD\,168112 actually result from the above scenario, then the properties of the shocks inside the wind deviate from those of strong shocks. This situation is very similar to the case of 9\,Sgr (Rauw et al.\,\cite{9sgr}) where the possible photon index was found to be larger than 2.9. The disagreement between our results and the hypothesis adopted by Chen \& White (\cite{CW}) could mean that the stellar winds harbour a distribution of hydrodynamical shocks characterized by different strengths (Feldmeier et al.\,\cite{Feld}). The assumption of strong shocks with $\chi$ = 4 only might therefore not hold in the case of a shock distribution. If the power law fitted as a second emission component in Sect.\,\ref{sect_fit} expresses any physical reality, the compression ratio deduced from our {\it XMM-Newton} observations should be considered as a mean value, typical of the shock distribution existing in the non-thermal X-ray emission region. At this stage, we emphasize nevertheless that the occurrence of a non-thermal emission process in the X-ray domain is not established. The possible reasons responsible for the non detection of inverse Compton scattering in X-rays are discussed in Sect.\,\ref{concl}. 

Although the variability in the X-ray domain is likely not compatible with the single star scenario, the radio variability could be explained by a variable magnetic field. Another possibility comes from shocks of different strengths crossing the relevant geometric region. The strongest shocks produce more radio emission than weaker ones, because they accelerate more of the electrons into the momentum range where synchrotron radiation is emitted at radio wavelengths (Van Loo et al.\,\cite{vanloo}).

\subsubsection{The binary scenario}
If HD\,168112 is a binary, in the case of a non-thermal hard X-ray emission, the discussion relevant to the photon index developed hereabove remains valid, but the shocks responsible for the electron acceleration are most probably associated with the wind-wind interaction (Williams et al.\,\cite{Wil3}; Eichler \& Usov\,\cite{EU}; Dougherty et al.\,\cite{DP}). The rather low value of the compression ratio derived in the previous section ($\chi$ = 1.75), compared to the strong shock value ($\chi_{str}$ = 4), could possibly be explained by heat conduction unopposed by magnetic field in the wind collision zone (Myasnikov \& Zhekov \cite{MZ}). Moreover, in a colliding wind binary, the non-thermal X-ray and radio emission would arise in the same region, i.e.\ they would be produced by the same population of relativistic electrons. In such a case, the determination of both radio (synchrotron) and X-ray (IC) luminosities could lead to a straightforward estimate of the magnetic field strength (see e.g.\ Benaglia \& Romero \cite{BR}).\\ 
We should note also that the strong variability of the non-thermal radio emitters is probably more compatible with binarity. Indeed, the observed variability for these objects can possibly be related to changes due to the orbital phase, like in WR\,140 (Williams et al.\,\cite{Wil2}).\\

For instance in an eccentric binary, seen under a low inclination, the strong variability of the radio emission could result from the synchrotron emission site, i.e. the wind interaction zone, moving in and out of the primary's free--free radio photosphere. Adopting the stellar parameters from Table\,\ref{param}, we estimate radii for the photospheres at 3.6, 6, and 20\,cm of 2.6, 3.8 and 8.8\,AU respectively. As discussed in Sect.\,\ref{ther}, the variations of the X-ray flux suggest an eccentricity of $\geq$ 0.5 for a putative binary system. The alternating detection and non-detection of HD\,168112 at 20\,cm (see Table\,\ref{table VLA flux results}) could be explained if the wind interaction zone lies outside the radio photosphere at apastron, whilst it would be buried within the photosphere at periastron passage. This translates into $$\mathrm{a}\,(1\,-\,\mathrm{e})\,\leq\,\mathrm{R}_{\tau\,=\,1}(20\,\mathrm{cm})\,\leq\,\mathrm{a}\,(1\,+\,\mathrm{e}).$$ If e = 0.5, one gets 5.9\,AU\,$\leq$\,a\,$\leq$\,17.6\,AU which corresponds to an orbital period 1.7\,yrs\,$\leq$\,P\,$\leq$\,8.9\,yrs for M$_1$ + M$_2$ = 70\,M$_\odot$, which is the stellar mass given by Leitherer (\cite{Leitherer88}).

\section{Conclusions \label{concl}}
The major points of our multiwavelength investigation of HD\,168112 can be summarized as follows:
\begin{enumerate}
\item[-] Our VLA data confirm the non-thermal nature of the radio emission from HD\,168112, and reveal also strong variability of the radio flux at 3.6 and 6\,cm. However, our target was not detected at 18 and 20\,cm.\\
\item[-] Our {\it XMM-Newton} data reveal what is most likely a thermal spectrum, with plasma temperatures up to 2 -- 3 keV. We failed to unambigously detect a high energy power law tail associated with a putative non-thermal component.\\
\item[-] The X-ray flux shows strong variability between our two {\it XMM-Newton} observations ($\sim$ 30 \%), mostly at energies above $\sim$\,1\,keV. If we assume that the X-ray and radio fluxes vary on time scales of months, our quasi-simultaneous {\it XMM-Newton} and VLA data suggest that the X-ray and radio fluxes vary in opposite directions: whereas the radio flux increased between March and September 2002, the X-ray flux had decreased.\\
\item[-] Our optical data failed to reveal a signature of binarity.
\end{enumerate}

Although the failure to detect radial velocity variations argues against HD\,168112 being a binary, it should be emphasized that many of the other properties of this system could be readily explained by a wind-wind interaction in a wide eccentric binary with a period of several years and seen under a low inclination angle. In fact, the radio and X-ray variations could result from the change in orbital separation between the two binary components. In this scenario, we could expect to measure the largest X-ray flux when the separation between the stars is minimum (i.e. around periastron), whereas the synchrotron radio emission would be strongest when the separation is larger, i.e. when the wind interaction zone lies well outside the radio photosphere of the primary star.\\
How could we establish the nature of HD\,168112? The best way to do this would probably be through a regular monitoring of the star in the radio domain. For instance, one radio observation every two months over $\sim$\,5 years might be sufficient to establish a recurrence in the radio light curve that might help us constrain the orbital period of a putative binary. A radio light curve with a quality comparable to that of WR\,140 (Williams et al.\,\cite{Wil2}; White \& Becker \cite{WB}) would undoubtedly provide new insight into the nature of this star. Another interesting test to be performed would be a high angular resolution optical interferometric observation of HD\,168112. At a distance of 2 kpc an interferometer with an angular resolution of $\sim$\,1 mas would be able to resolve a secondary component if it lies at a distance of at least $\sim$\,2 AU from the primary. This technique was succesfully recently applied to the case of two spectroscopic binaries (Monnier et al.\,\cite{Monnier}).\\

Finally, it is important to ask the following question: what could be the reason for our failure to detect an inverse Compton X-ray emission although we knew all the ingredients required to produce such an emission to be present in HD\,168112 and 9\,Sgr (Rauw et al.\,\cite{9sgr})? Several factors could contribute to hide or inhibit such an inverse Compton X-ray emission:
\begin{enumerate}
\item[1.] {\it An overwhelming thermal X-ray emission.} This could be an issue especially for binary systems where the interaction region could produce a fair amount of hard thermal X-rays. In this respect, observations in the $\gamma$-ray energy range will be extremely useful. A detection of early-type stars with the {\it INTEGRAL} observatory would not be affected by the thermal emission from the colliding wind region and would provide an unambigous evidence for the inverse Compton mechanism.\\
\item[2.] {\it Too large a dilution of the photospheric UV flux at the site of the relativistic electron acceleration.} Again, this is more likely to be an issue in a binary system with a wide orbital separation where the electron acceleration takes place in the wind interaction zone, i.e. several tens or hundreds of stellar radii away from the UV photosphere.\\
\item[3.] {\it Too large a magnetic field.} If the non-thermal X-ray and radio emission arise from the same population of relativistic electrons, the ratio between the IC and synchrotron luminosities should depend on the magnetic field strengths: $$ \frac{\mathrm{L}_\mathrm{sync}}{\mathrm{L}_\mathrm{IC}}\,\propto\,\frac{\mathrm{B}^2\,\mathrm{d}^2}{\mathrm{L}}$$ where L$_\mathrm{sync}$, L$_\mathrm{IC,}$, B, d, and L are respectively the radio synchrotron luminosity, the inverse Compton luminosity, the strength of the magnetic field, the distance to the star, and its bolometric luminosity (Benaglia \& Romero\,\cite{BR}). Therefore, for a rather large value of B, the synchrotron luminosity could be large, whereas the IC X-ray emission would remain undetectable.
\end{enumerate}

Clearly more work is needed on the theoretical side to provide simulated inverse Compton X-ray and $\gamma$-ray spectra of single and binary early-type stars suitable for a comparison with the new data that are now available.

\acknowledgement{Our thanks go to Alain Detal (Li\`ege) for his help in 
installing the {\sc sas}. The authors wish also to thank Dr. Eric Gosset for instructive discussions, as well as Prof. Rolf Mewe for very helpful instructions concerning the transition rate calculations. The Li\`ege team acknowledges support from the Fonds National de la Recherche Scientifique (Belgium) and through the PRODEX XMM-OM and Integral Projects. This research is also supported in part by contracts P4/05 and P5/36 ``P\^ole d'Attraction Interuniversitaire'' (SSTC-Belgium). JMP gratefully acknowledges funding from PPARC for a PDRA position. SVL gratefully acknowledges a doctoral research grant by the Belgian State, Federal Office for Scientific, Technical and Cultural 
Affairs (OSTC). This research has made use of the SIMBAD database, operated at CDS, Strasbourg, France and NASA's Astrophysics Data System Abstract Service.}

\end{document}